\pdfoutput = 1
\documentclass[12pt, draftclsnofoot, onecolumn]{IEEEtran}

\usepackage{cite}
\usepackage{amsmath}
\usepackage{amssymb}
\usepackage{array}

\ifCLASSOPTIONcompsoc
\usepackage[caption=false,font=normalsize,labelfont=sf,textfont=sf]{subfig}
\else
\usepackage[caption=false,font=footnotesize]{subfig}
\fi

\usepackage{url}
\usepackage{nth}
\usepackage{bm}
\usepackage[export]{adjustbox}
\usepackage{siunitx}
\DeclareSIUnit\dBm{dBm}

\usepackage{tikz}
\usepackage{pgfplots}
\pgfplotsset{compat=newest}
\usepgfplotslibrary{colormaps}
\usetikzlibrary{backgrounds, calc, decorations, decorations.markings, decorations.pathreplacing, math, arrows.meta, shapes}
\pgfdeclarelayer{bg}    
\pgfsetlayers{bg,main}  

\makeatletter
\renewcommand*\env@matrix[1][c]{\hskip -\arraycolsep
	\let\@ifnextchar\new@ifnextchar
	\array{*\c@MaxMatrixCols #1}}
\makeatother

\usepackage{array}
\usepackage{multirow}
\usepackage{mathtools}
\usepackage{ulem}
\usepackage{todonotes}

\newcommand{\E}[2]{\mathbb{E}_{#2}\!\left[#1\right]}
\newcommand{\vect}[1]{\bm{#1}}
\newcommand{\trans}{^{\mathrm{T}}}
\newcommand{\herm}{^{\mathrm{H}}}
\newcommand{\distas}[1]{\overset{#1}{\sim}}%
\newcommand{\e}{\mathrm{e}}

\DeclareMathOperator{\jj}{j}

\DeclareMathOperator*{\trace}{tr}

\DeclareMathOperator*{\atan2}{atan2}

\newtheorem{theorem}{Theorem}
\newtheorem{definition}{Definition}

\usepackage{textcomp}

\usepackage[pdfencoding=auto, psdextra]{hyperref}
\usepackage{bookmark}

\title{Performance Limits of Single-Anchor Millimeter-Wave Positioning}

\author{Anastasios~Kakkavas,~\IEEEmembership{Student Member,~IEEE,}
	Mario~H.~Casta\~neda~Garcia,
	Richard~A.~Stirling-Gallacher,
	and~Josef~A.~Nossek,~\IEEEmembership{Life~Fellow,~IEEE}%
	
	\thanks{A.~Kakkavas, M.~H.~Casta\~{n}eda~Garcia and R.~A.~Stirling-Gallacher are with the Munich Research Center, Huawei Technologies Duesseldorf GmbH, 80992 Munich, Germany (e-mail: \{anastasios.kakkavas, mario.castaneda, richard.sg\}@huawei.com). A.~Kakkavas and J.~A.~Nossek are with the Department of Electrical and Computer Engineering, Technical University of Munich, 80333 Munich, Germany (e-mail: josef.a.nossek@tum.de). J.~A.~Nossek is also with the Department of Teleinformatics Engineering, Federal University of Ceara, 60020-180 Fortaleza, Brazil}
}

\hypersetup{
	pdfinfo={
		Title={Performance Limits of Single-Anchor Millimeter-Wave Positioning},
		Author={Anastasios Kakkavas, Mario H. Casta\~neda Garcia, Richard A. Stirling-Gallacher, Josef A. Nossek}
	}
}

\usepackage{textcomp}

\usepackage{eso-pic}

\begin{document}
	\maketitle
	\tikz[overlay,remember picture]
	{
		\node at ($(current page.north west)+(1in,-0.3cm)$) [rotate=0, anchor=north west] {\parbox{\textwidth - 0.3in}{\footnotesize \footnotesize A. Kakkavas, M. H. Casta\~neda Garc\'ia, R. A. Stirling-Gallacher and J. A. Nossek, "Performance Limits of Single-Anchor Millimeter-Wave Positioning," in IEEE Transactions on Wireless Communications, vol. 18, no. 11, pp. 5196-5210, Nov. 2019, doi: 10.1109/TWC.2019.2934460.}};
		
		\node at ($(current page.south west)+(1in,0.3cm)$) [rotate=0, anchor=south west] {\parbox{\textwidth}{\footnotesize \footnotesize 2019 IEEE. Personal use of this material is permitted. Permission from IEEE must be obtained for all other uses, in any current or future media, including reprinting/republishing this material for advertising or promotional purposes, creating new collective works, for resale or redistribution to servers or lists, or reuse of any copyrighted component of this work in other works.}};
	}

	\normalem 

	\vspace*{-0.35cm}
	\begin{abstract}
		The fundamental limits of single-anchor multi-antenna positioning are investigated. Exploiting the structure of the multiple input-multiple output-orthogonal frequency division multiplexing (MIMO-OFDM) channel at millimeter-wave frequencies, we present geometrically intuitive asymptotic expressions for the Fisher information on position, orientation and velocity for large bandwidth and number of antennas. The effects of synchronization errors and mobility are studied and it is shown that 
		non-line-of-sight (NLOS) paths can be used to estimate the synchronization error and drastically improve the positioning performance. We also find that, in the presence of line-of-sight (LOS), mobility has a small impact on the achievable positioning accuracy, but in the NLOS-only scenario it can significantly improve the achievable performance, depending on the variance of the synchronization error. 
		Finally, considering a communication system with device-specific transmission and reception constraints, we compare the positioning accuracy between the downlink and the uplink and show that they are equivalent under the same signal-to-noise ratio (SNR).
	\end{abstract}
	
	
	\IEEEpeerreviewmaketitle
	
	\section{Introduction}
		\label{sec:Introduction}
		\IEEEPARstart{T}{he} \nth{5} generation (5G) of wireless communication networks is on its way, with the vision of providing unprecedentedly high data rates, massive connectivity and reduced latency, among others, which can be enabled by the large bandwidth available at millimeter-wave (mm-Wave) frequencies, massive number of antennas and dense deployment of base stations. Coincidentally or not, these enablers provide fertile ground for a radical improvement of the positioning capabilities of wireless communication networks~\cite{DRZ15} and bring localization to the foreground of 5G development. Apart from fusion with Global Navigation Satellite Systems (GNSS) to enhance traditional positioning services, the expected improvement of localization accuracy of communication networks can be a key driving force in the realization of scenarios such as assisted living~\cite{WML+16}, smart factories~\cite{WHK+16} and automated driving~\cite{WSD+17,KCS+18}, where GNSS and other existing technologies may not be able to guarantee the desired positioning accuracy under all conditions. Moreover, it is anticipated that position information will be used as an input to communication-related tasks, such as proactive resource allocation~\cite{HWC+15}, beamforming~\cite{KCT+16} and beam-alignment~\cite{MGK+17}.
		
		In an effort to unlock the great potential of next generation wireless communication networks, numerous works on the topic have been published. Theoretical analyses for positioning under a multi-anchor setup have been presented in~\cite{SW10,SW10B,SW07,SW09, HSZ+16}. Despite promising high accuracy, multi-anchor positioning might be impossible at mm-Wave due to the lack of significantly strong links to two or more anchors. In mm-Wave frequencies the path loss between isotropic antennas\footnote{Isotropic antennas are only assumed as a theoretical tool; in practice, antenna elements with potentially similar, but not identical, properties are used.} is increased as a consequence of their decreased effective area/aperture and beamforming gains are required to compensate for it. 
		On the other hand, the fact that more antennas can be packed in the same physical area allows for multi-antenna transceivers, which enable accurate angle of departure (AOD) and angle of arrival (AOA) estimation. Also, the large bandwidth available at mm-Wave frequencies enables accurate time of arrival (TOA) or time difference of arrival (TDOA) measurements. Hence, reliable position estimation is possible with a single anchor.
		
		In recent years, many studies on single-anchor positioning have been published. The single-anchor localization bounds were derived in~\cite{GGD15} for arrays with different beamforming strategies, under line-of-sight (LOS)-only propagation, taking synchronization errors and beamforming weights' quantization errors into account. This work was extended in \cite{GGD17} including MIMO transmission and multipath propagation. In~\cite{SGD+15} the Cram\'{e}r-Rao lower bound (CRLB) for single-anchor LOS positioning was presented, deriving a necessary condition on the reference signal for a non-singular Fisher information matrix (FIM). The authors of~\cite{AZA+18} provided asymptotic expressions for the position error bounds in sparse mm-Wave channels with large bandwidth and large number of antennas, and conducted a comparison between downlink and uplink positioning. 
		In~\cite{MWB+17} it was shown that the FIM provided by single-bounce NLOS paths is rank-$1$ and analytic expressions for the direction and intensity of position and orientation information were obtained. In~\cite{SGD+18}, the single-anchor localization error bounds with a multicarrier waveform were derived and an algorithm approaching these bounds for a high SNR was presented. Considering general non-orthogonal waveforms and TOA and AOA-based positioning, the authors of \cite{WWS19} showed analytically that multipath effects vanish in the large array regime and developed a novel bound-achieving direct localization algorithm. Single-anchor positioning algorithms have also been presented in~\cite{TVD+17} and~\cite{MWB18}. Apart from \cite{GGD15} and~\cite{GGD17}, the aforementioned works assume no synchronization errors, which is difficult to obtain in practical communication systems. To address this issue, the authors of~\cite{AWA+18} propose a two-way localization protocol. In \cite{LSW18} a joint localization and synchronization algorithm is presented for a network with a single anchor and multiple full-duplex agents.
		
		Following previous works~\cite{AZA+18,MWB+17,SGD+18}, we study the performance limits of two-dimensional (2D) single-anchor multi-antenna positioning in sparse mm-Wave channels, where only single-bounce NLOS paths are assumed strong enough for reception~\cite{VB03,RBM+12,MGP+14,MMMD21}. We extend prior works by considering three aspects: a multicarrier waveform\footnote{The results are also applicable for the single-carrier frequency division multiple access (SC-FDMA) with appropriate design of the reference signal.}, the impact of synchronization errors and mobility. To this end, we provide an analysis of the Fisher information associated with position, orientation, and velocity estimation in mm-Wave MIMO systems. The main contributions of this work are the following:
		\begin{itemize}
			\item When both the transmitter (Tx) and receiver (Rx) are static (static scenario), we show that the directions of information provided by the single-bounce NLOS paths depend only on the geometry of the channel and not on the system parameters. 
			\item When the receiver or the transmitter is moving (dynamic scenario), we study the effect of the Doppler shift on the Fisher information. We prove that the information provided by each single-bounce NLOS path is rank-$2$. 
			\item For the static and dynamic scenario, the impact of the synchronization error is analytically investigated and numerical evaluations of its effect on positioning with or without the LOS path are provided.
			\item We compare donwlink (DL) and uplink (UL) positioning under device-specific Tx/Rx constraints and show that the link with the highest receive SNR offers the highest accuracy. For perfectly synchronized and asynchronous systems we show that the FIMs of the position and orientation parameters in the DL and the UL differ only by a scalar, which is equal to the ratio of the receive SNR in the DL and the UL.
		\end{itemize}
		
		The rest of the paper is organized as follows. The system model is derived in Section II. In Section III, the Cram\'{e}r-Rao bound for position and orientation estimation in the static and dynamic scenario is derived. The relation between DL and UL positioning is derived in Section IV. The asymptotic analysis and geometric interpretation of the Fisher information are provided in Section V. Numerical evaluations of the bounds are provided in Section VI. Section VII concludes the work.
		
		\textbf{Notation:} We use bold lowercase for column vectors, bold uppercase for matrices, non-bold for scalars and calligraphic letters for sets. Depending on its argument, $|\cdot|$ denotes the absolute value of a scalar, the determinant of a matrix or the cardinality of a set. The operators $\left(\cdot\right)\trans$ and $\left(\cdot\right)\herm$ denote the transpose and the conjugate transpose of a vector or matrix, $\|\cdot\|_2$ denotes the Euclidean norm of a vector and $\Re\left\{\cdot\right\}$ and $\Im\left\{\cdot\right\}$ denote the real and imaginary part of a complex number. The $i$-th element of a vector and the $(i,j)$-th element of a matrix are denoted by $[\cdot]_i$ and $[\cdot]_{i,j}$, respectively. The expectation operator is denoted by $\E{\cdot}{}$ and the sets of real and complex numbers are denoted by $\mathbb{R}$ and $\mathbb{C}$. A multivariate circularly symmetric complex Gaussian distribution with mean $\vect{\mu}$ and covariance matrix $\vect{C}$ is denoted by $\mathcal{N}_{\mathbb{C}}\left(\vect{\mu}, \vect{C}\right)$.
		A list of frequently used symbols is provided in Table~\ref{tab:list of frequently used symbols}.
		\begin{table}
			\centering
			\caption{List of frequently used symbols}
			\begin{tabular}{p{2cm} p{5.5cm}} 
				$\vect{p}_{\text{T}}$, $\alpha_{\text{T}}$, $\vect{v}_{\text{T}}$ & Tx position, orientation and velocity\\
				$\vect{p}_{\text{R}}$, $\alpha_{\text{R}}$, $\vect{v}_{\text{R}}$ & Rx position, orientation and velocity\\
				$\vect{p}_{\text{s},l}$ & position of $l$-th scatterer/reflector\\
				$d_{\text{T}, \text{R}}$ & Tx-Rx distance\\
				$d_{\text{T},\text{s},l}$, $d_{\text{R},\text{s},l}$ & distance of the $l$-th scatterer/reflector from the Tx and the Rx\\
				$\vect{u}(\theta)$, $\vect{u}_{\perp}(\theta)$ & unit vectors at angles $\theta$ and $\theta - \pi/2$\\
				$N_{\text{T}}$, $M_{\text{T}}$ & number of Tx antennas and beamforming vectors\\
				$N_{\text{R}}$, $M_{\text{R}}$ & number of Rx antennas and beamforming vectors\\
				$S_{\text{T}}(\tilde{\theta})$, $S_{\text{R}}(\tilde{\theta})$ & Tx and Rx squared array aperture function\\
				$L$ & number of propagation paths\\
				$\tau_l$, $\theta_{\text{T},l}$, $\theta_{\text{R},l}$, $h_l$ & TOA, AOD, AOA and gain of path $l$\\
				$\xi_l$ & intensity effect of Doppler shift on path $l$\\
			\end{tabular}
			\begin{tabular}{p{1.7cm} p{5.5cm}}
				$t_{\text{rms},l}$ & root mean square duration of the signal propagated through path $l$\\
				$\rho_l$ & transverse velocity component of path $l$\\
				$N$, $\mathcal{P}$, $N_{\text{B}}$ & number of subcarriers, set of used subcarriers and number of OFDM symbols\\
				$F_{\text{s}}$, $f_{\text{c}}$, $c$ & sampling rate, carrier frequency and speed of light\\
				$P_{\text{T}}$, $\gamma_p$ & Tx power and its fraction allocated to subcarrier $p$\\
				$\beta$ & effective baseband bandwidth\\
				$\bar{\omega}_c$ & effective carrier angular frequency\\
				$\epsilon_{\text{clk}}$, $\sigma_{\text{clk}}^2$ & clock synchronization error and its variance\\
				$\sigma_{\eta,\text{R}}^2$ & noise variance per dimension\\
				$\delta_{\text{R},\text{T}}$ & receive SNR\\
				
			\end{tabular}
			\label{tab:list of frequently used symbols}
		\end{table}

	\section{System Model}
		\label{sec:system model}
		In this section we present the signal model when both the transmitter and the receiver are static, and then we extend the model to the dynamic scenario, where either the transmitter or the receiver is moving.
		
		\subsection{Static Scenario} 
		\subsubsection{Geometry}
\begin{figure}
			\centering
			\includegraphics[scale=1]{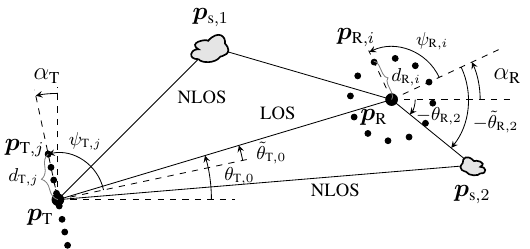}
			\caption{Geometric model, example with a uniform linear array (ULA) at the transmitter and a uniform circular array (UCA) at the receiver.}
			\label{fig:geometric_model}
		\end{figure}

		In Fig.~\ref{fig:geometric_model} an example of the considered 2D setup is shown.
		The Tx and Rx have $N_{\text{T}}$ and $N_{\text{R}}$ antennas with reference points $\vect{p}_{\text{T}} = \left[p_{\text{T,x}},\; p_{\text{T,y}}\right]\trans\in\mathbb{R}^2$ and $\vect{p}_{\text{R}} = \left[p_{\text{R,x}},\; p_{\text{R,y}}\right]\trans\in\mathbb{R}^2$. Their orientation with respect to (w.r.t.) the horizontal line is $\alpha_{\text{T}}$ and $\alpha_{\text{R}}$, respectively. The position of the $j$-th element of the Tx array is
		\begin{IEEEeqnarray}{rCl}
			\vect{p}_{\text{T},j} &=& \vect{p}_{\text{T}} + d_{\text{T},j}\vect{u}\left(\psi_{\text{T},j} + \alpha_{\text{T}}\right) = \left[p_{\text{T},j,\text{x}},\; p_{\text{T},j,\text{y}}\right]\trans\in\mathbb{R}^2, \quad j=1,\dots, N_{\text{T}},
			\IEEEeqnarraynumspace 
		\end{IEEEeqnarray}
		where $\vect{u}(\psi) = \left[\cos(\psi), \; \sin(\psi)\right]\trans$ and $d_{\text{T},j}$ and $\psi_{\text{T},j} + \alpha_{\text{T}}$ are its distance and angle from the Tx reference point, as shown in Fig.~\ref{fig:geometric_model}. The quantities for the Rx array are defined accordingly. 
		
		We assume that for all the antenna pairs there are $L$ discrete propagation paths. The first path ($l=0$) is either the LOS path, if it exists, or a single-bounce NLOS path, while the rest $L-1$ are single-bounce NLOS paths ($l=1,\ldots,L-1$). The point of incidence of the $l$-th single-bounce path, which corresponds either to scattering or reflection, is $\vect{p}_{\text{s},l} = \left[p_{\text{s},l,\text{x}},\; p_{\text{s},l,\text{y}}\right]\trans$. The array apertures are assumed to be small compared to the distance between Tx and Rx, as well as the distance between each of the scatterers/reflectors and the transmitter and receiver. Therefore, the delay of path $l$ from Tx element $j$ to Rx element $i$ can be approximated by $\tau_{l,i,j} \approx \tau_l - \tau_{\text{T},j}(\tilde{\theta}_{\text{T},l}) - \tau_{\text{R},i}(\tilde{\theta}_{\text{R},l}), \; l=0, \ldots, L-1,$ where  
		\begin{IEEEeqnarray}{rCl}
			\tau_l &=& \begin{cases} 
				\left\|\vect{p}_{\text{R}} - \vect{p}_{\text{T}}\right\|_2/c, & \text{if the } l\text{-th path is LOS},\\
				\left(\left\|\vect{p}_{\text{s},l} - \vect{p}_{\text{T}}\right\|_2 + \left\|\vect{p}_{\text{R}} - \vect{p}_{\text{s},l}\right\|_2\right)/c, & \text{otherwise},
			\end{cases}
		\end{IEEEeqnarray}
		$\tau_{\text{T},j}(\tilde{\theta}_{\text{T},l}) = d_{\text{T},j}\vect{u}\trans\left(\psi_{\text{T},j}\right)\vect{u}(\tilde{\theta}_{\text{T},l})/c$, and  $\tau_{\text{R},i}(\tilde{\theta}_{\text{R},l}) =  d_{\text{R},i}\vect{u}\trans\left(\psi_{\text{R},i}\right)\vect{u}(\tilde{\theta}_{\text{R},l})/c$, with $c$ being the speed of light. We assume without loss of generality that $\tau_0\leq\tau_1\leq\ldots\leq\tau_{L-1}$. The angles are defined as $\tilde{\theta}_{\text{T},l} = \theta_{\text{T},l} - \alpha_{\text{T}},\;\tilde{\theta}_{\text{R},l} = \theta_{\text{R},l} - \alpha_{\text{R}}, \; l=0,\ldots,L-1$, where $\theta_{\text{T},0} = \atan2\left(p_{\text{R},\text{y}} - p_{\text{T}, \text{y}}, p_{\text{R},\text{x}} - p_{\text{T}, \text{x}}\right), \theta_{\text{R},0} = \theta_{\text{T},0} + \pi, \theta_{\text{T},l} = \atan2\left(p_{\text{s},l,\text{y}} - p_{\text{T}, \text{y}}, p_{\text{s},l,\text{x}} - p_{\text{T}, \text{x}}\right) - \alpha_{\text{T}}$ and $\theta_{\text{R},l} = \atan2\left(p_{\text{s},l,\text{y}} - p_{\text{R}, \text{y}}, p_{\text{s},l,\text{x}} - p_{\text{R}, \text{x}}\right) - \alpha_{\text{R}}, \; l=1,\ldots,L-1$, with $\atan2\left(y,x\right)$ being the four-quadrant inverse tangent function.
		
		\subsubsection{Signal Model}
		Assuming high resolution analog-to-digital conversion (ADC) and digital-to-analog conversion (DAC), such that the system is practically linear, the pulse $\tilde{g}(t)$ models all the Tx and Rx processing steps, namely upsampling, digital Tx filtering/pulse shaping, DAC, analog Tx and Rx filtering, ADC, digital Rx filtering/pulse shaping and downsampling.
		Hence, the discrete-time channel impulse response between Tx antenna $j$ and Rx antenna $i$ is 
		\begin{IEEEeqnarray}{rCl}
			c_{i,j}[m] &=& \sum_{l=0}^{L-1} h_l'\tilde{g}\left(mT_{\text{s}} - \tau_{l,i,j} \right)\e^{-\jj 2\pi f_{\text{c}} \tau_{l,i,j}},
			\IEEEeqnarraynumspace
		\end{IEEEeqnarray}
		where $T_{\text{s}}$ is the sampling period, $F_{\text{s}} = 1/T_{\text{s}}$ is the sampling rate and $h_l'\in\mathbb{C},\;l=0,\ldots, L-1,$ are the gains of the propagation paths. 
		Then, the received signal at antenna $i$ for sample $n$ is
		\begin{IEEEeqnarray}{rCl}
			\left[\tilde{\vect{r}}[n]\right]_i &=& \sum_{j=1}^{N_{\text{T}}} \sum_{m=-\infty}^{\infty} c_{i,j}[m] \left[\tilde{\vect{x}}\left[n-m\right]\right]_j + \left[\tilde{\vect{\eta}}[n]\right]_i,
			\IEEEeqnarraynumspace
		\end{IEEEeqnarray}
		where $\tilde{\vect{x}}\left[n\right]\in\mathbb{C}^{N_{\text{T}}}$ is the time-domain transmit signal at sample $n$ and $\tilde{\vect{\eta}}[n]$ is the zero-mean circularly-symmetric Gaussian noise vector, which is temporally and spatially uncorrelated with variance $2\sigma_{\eta,\text{R}}^2$ per-antenna, i.e. $\tilde{\vect{\eta}}[n] \distas{} \mathcal{N}_{\mathbb{C}}(\vect{0}, 2\sigma_{\eta,\text{R}}^2\vect{I}_{N_{\text{R}}})$.
		
		We consider a Cyclic Prefix-OFDM (CP-OFDM) system with $N$ subcarriers and the CP consisting of $N_{\text{CP}}$ samples. $N_{\text{B}}$ OFDM symbols are transmitted and $\mathcal{P}$ is the set of indices of the loaded subcarriers. In order to ensure there is no inter-symbol interference, the CP is assumed to be long enough to accomodate the effective delay spread of the channel, including the length of the pulse $\tilde{g}(t)$ and the frame synchronization error. In mm-Wave systems the dimensionality of the digital transmit and receive signals may be reduced by hardware constraints, i.e. the transmitter and the receiver may be equipped with $M_{\text{T}}\leq N_{\text{T}}$ and $M_{\text{R}}\leq N_{\text{R}}$ radio frequency (RF) chains, respectively. Therefore, we consider a transmit precoder $\vect{F}_{\text{T}}\in\mathbb{C}^{N_{\text{T}}\times M_{\text{T}}}$ and a receive combiner $\vect{F}_{\text{R}}\in\mathbb{C}^{N_{\text{R}}\times M_{\text{R}}}$ that are common to all subcarriers. When no precoding (receive combining) is applied $N_{\text{T}} = M_{\text{T}}$ ($N_{\text{R}} = M_{\text{R}}$) and $\vect{F}_{\text{T}} = \vect{I}_{N_{\text{T}}}$ ($\vect{F}_{\text{R}} = \vect{I}_{N_{\text{R}}}$).
		The Tx signal vector at subcarrier $p$, $p\in\mathcal{P}$, of symbol $b$, $b=0,\ldots, N_{\text{B}} - 1$, before precoding is ${\vect{s}}_b[p]\in\mathbb{C}^{M_{\text{T}}}$, and after precoding is $\vect{x}_b\left[p\right] = \vect{F}_{\text{T}}^*{\vect{s}}_b\left[p\right]$. Without loss of generality, we assume that the reference signal transmission starts at time $0$. The time-domain transmit signal (after precoding) $\tilde{\vect{x}}\left[n\right]$ is related to the $b$-th time-domain OFDM symbol $\tilde{\vect{x}}_b[m]$ as
		$\tilde{\vect{x}}\left[bM + m\right] = \tilde{\vect{x}}_b[m]$, where $M = N + N_{\text{CP}}$ and $\tilde{\vect{x}}_b[m] = \sum_{p\in\mathcal{P}}  {\vect{x}}_b\left[p\right] \e^{\jj \frac{2\pi}{N}p \left(m - N_{\text{CP}}\right)}/\sqrt{N}, \;m=0\ldots,M-1$.	
		
		We assume that the receiver is imperfectly synchronized, it namely works under the premise that the signal was transmitted at $\epsilon_{\text{clk}}$, where $\epsilon_{\text{clk}}$ is the clock synchronization error, which we assume to be a zero-mean Gaussian distributed random variable with variance $\sigma_{\text{clk}}^2$. We note that $\sigma_{\text{clk}}^2=0$ corresponds to perfectly synchronized clocks and $\sigma_{\text{clk}}^2 \rightarrow \infty$ corresponds to asynchronous operation.
		After receive combining and Discrete Fourier Transform (DFT), the receiver uses its knowledge of the time of transmission to apply an appropriate per-subcarrier rotation to the post-DFT signal. It can be shown that the resulting signal vector at subcarrier $p$ of symbol $b$ is
		\begin{IEEEeqnarray}{rCl}
			\vect{w}_b[p] &\approx&  \vect{F}_{\text{R}}\herm\vect{H}[p] \vect{F}_{\text{T}}^*\vect{s}_b[q] + \vect{F}_{\text{R}}\herm\vect{\eta}_b[p],
			\label{eq:signal model}
			\IEEEeqnarraynumspace
		\end{IEEEeqnarray}
		where
		\begin{IEEEeqnarray}{rCl}
			\vect{H}[p] &=& {g}[p] \e^{-\jj \omega_p \epsilon_{\text{clk}}} \sum_{l=0}^{L-1}  h_l  \e^{-\jj \omega_p\tau_l}\vect{a}_{\text{R},p}(\tilde{\theta}_{\text{R},l}) \vect{a}_{\text{T},p}\trans(\tilde{\theta}_{\text{T},l})\in\mathbb{C}^{N_{\text{R}}\times N_{\text{T}}},\label{eq:static channel definition}\\
			\vect{a}_{\text{T},p}(\tilde{\theta}_{\text{T},l}) &=& \begin{bmatrix}
				e^{\jj \left(\omega_c+\omega_p\right) \tau_{\text{T},1}(\tilde{\theta}_{\text{T},l})}, & \ldots, &\e^{\jj \left(\omega_c+\omega_p\right) \tau_{\text{T},N_{\text{T}}}(\tilde{\theta}_{\text{T},l})}
			\end{bmatrix}\trans\in\mathbb{C}^{N_{\text{T}}},
			\IEEEeqnarraynumspace
		\end{IEEEeqnarray}
		with $h_l = h_l'\e^{-\jj 2\pi f_{\text{c}} \tau_l}$, $\omega_p = 2 \pi p F_{\text{s}}/N$, $\omega_c = 2\pi f_{\text{c}}$ and $\vect{\eta}_b[p] \distas{} \mathcal{N}_{\mathbb{C}}(\vect{0}, 2\sigma_{\eta,\text{R}}^2\vect{I}_{N_{\text{R}}})$. $\vect{a}_{\text{R},p}(\tilde{\theta}_{\text{R},l})$ is defined similarly to $\vect{a}_{\text{T},p}(\tilde{\theta}_{\text{T},l})$. 
		We assume that the range of the angles passed as arguments to the array steering vectors is such that, for a given array structure, there is 1-1 mapping between angles and steering vectors. The approximate equality in \eqref{eq:signal model} follows from the assumption that, apart from being time-limited, $\tilde{g}(t)$ is essentially band-limited.
		
		\subsection{Dynamic Scenario}
		We now extend the signal model to the dynamic scenario, with the receiver moving with constant velocity 
		$\vect{v}_{\text{R}} = \left[v_{\text{R,x}}, \; v_{\text{R,y}}\right]\trans\in\mathbb{R}^2$. Let $n_0T_{\text{s}}$ be the TOA of the first sample of the received signal fed to the DFT and $\vect{p}_{\text{R},0}$ be the receiver's position at that time instant. The position of the receiver at time $nT_{\text{s}}$ is $\vect{p}_{\text{R}}(nT_{\text{s}}) = \vect{p}_{\text{R},0} + \vect{v}_{\text{R}}\cdot\left(n - n_{0}\right)T_{\text{s}}$. As in the static scenario, we make use of the far field assumption, so that the angles and the channel gains are approximately invariant during the observation interval $N_{\text{B}} M T_{\text{s}}$. The received signal at antenna $i$ is
		\begin{IEEEeqnarray}{rCl}
			\left[\tilde{\vect{r}}[n]\right]_i &=& \sum_{j=1}^{N_{\text{T}}} \sum_{m=-\infty}^{\infty} c_{i,j}[m, n] \left[\tilde{\vect{x}}\left[n-m\right]\right]_j + \left[\tilde{\vect{\eta}}[n]\right]_i,
			\IEEEeqnarraynumspace
		\end{IEEEeqnarray}
		where
		\begin{IEEEeqnarray}{rCl}
			c_{i,j}[m,n] &=& \sum_{l=0}^{L-1} h_l'\tilde{g}\left(mT_{\text{s}} - \tau_{l,i,j} + v_l\left(n - n_{0}\right)T_{\text{s}}/c\right)\e^{-\jj 2\pi f_{\text{c}} \left(\tau_{l,i,j} - v_l\left(n - n_{0}\right)T_{\text{s}}/c\right)},
			\IEEEeqnarraynumspace
		\end{IEEEeqnarray}
		with $v_{l} = \vect{v}_{\text{R}}\trans \vect{u}\left(\theta_{\text{R},l}\right)$ being the speed of the receiver in the direction of the AOA.
		We can show that the signal vector at subcarrier $p$ of symbol $b$ after receive combining is
		\begin{IEEEeqnarray}{rCl}
			\vect{w}_b[p] &\approx&  \sum_{q\in\mathcal{P}}\vect{F}_{\text{R}}\herm\vect{H}_{b}[p,q] \vect{F}_{\text{T}}^*\vect{s}_b[q] + \vect{F}_{\text{R}}\herm\vect{\eta}_b[p],
			\label{eq:signal model dynamic}
			\IEEEeqnarraynumspace
		\end{IEEEeqnarray}
		where
		\begin{IEEEeqnarray}{rCl}
			\vect{H}_{b}[p,q] &=& g[q] \e^{-\jj \omega_q\epsilon_{\text{clk}}}\sum_{l=0}^{L-1} h_l  \e^{-\jj \left(\omega_q\tau_l - \frac{\omega_c + \omega_q}{c}v_{l} bMT_{\text{s}}\right)} Q\left(\left( \omega_p - \omega_q - \frac{\omega_c + \omega_q}{c}v_{l} \right)\frac{T_{\text{s}}}{2}\right)\times\nonumber\\
			&&\vect{a}_{\text{R},q}(\tilde{\theta}_{\text{R},l}) \vect{a}_{\text{T},q}\trans(\tilde{\theta}_{\text{T},l})\in\mathbb{C}^{N_{\text{R}}\times N_{\text{T}}},\label{eq:dynamic channel definition}
			\IEEEeqnarraynumspace
		\end{IEEEeqnarray}
		with $Q(x) = \e^{-\jj (N - 1) x} \sin(Nx)/(N\sin(x)).$
		The Doppler shift $v_{l}\left(\omega_c + \omega_q\right)/c$ is subcarrier-dependent and an increasing function of the subcarrier index. Similar to \cite{HSZ+16}, we assume that the observation interval is much smaller than the channel coherence time $T_{\text{co}} \propto c/\left(f_{\text{c}} \|\vect{v}_{\text{R}}\|_2\right)$~\cite{Rap01}. When the transmitter is moving with velocity $\vect{v}_{\text{R}}$ and the receiver is static, the signal model \eqref{eq:signal model dynamic} is still valid with $v_l = \vect{v}_{\text{T}}\trans \vect{u}\left(\theta_{\text{T},l}\right)$.

	\section{Computation of the Cram\'{e}r-Rao Lower Bound}
		\label{sec:CRLB computation}
		\subsection{Static Scenario}
		Based on the received signal model \eqref{eq:signal model}, we now derive the Fisher information matrix and the Cram\'er-Rao lower bound for position and orientation estimation for a single-anchor MIMO-OFDM system. Setting $h_{l,\Re} = \Re\{h_{l}\}$ and $h_{l,\Im} = \Im\{h_{l}\}, \; l=0,\ldots,L-1,$ we define the channel parameter vector $\vect{\phi}$ as
		\begin{IEEEeqnarray}{rCl}
			\vect{\phi} &=& [\epsilon_{\text{clk}}, \tau_0, \tilde{\theta}_{\text{T},0}, \tilde{\theta}_{\text{R},0}, h_{0, \Re}, h_{0, \Im}, \ldots, \tau_{L-1}, \tilde{\theta}_{\text{T},{L-1}}, \tilde{\theta}_{\text{R},{L-1}}, h_{{L-1},\Re}, h_{{L-1},\Im}
			]\trans \in \mathbb{R}^{5L + 1}.\label{eq:channel parameter vector}
			\IEEEeqnarraynumspace
		\end{IEEEeqnarray}
		The covariance matrix $\vect{C}_{\hat{\vect{\phi}}}$ of any unbiased estimator $\hat{\vect{\phi}}$ satisfies~\cite{VT13} $\vect{C}_{\hat{\vect{\phi}}} - \vect{J}_{\vect{\phi}}^{-1} \succeq \vect{0}$, where $\succeq \vect{0}$ denotes positive semi-definiteness and $\vect{J}_{\vect{\phi}}\in\mathbb{R}^{(5L+1)\times (5L+1)}$ is the hybrid FIM of $\vect{\phi}$. $\vect{J}_{\vect{\phi}}$ is defined as
		\begin{IEEEeqnarray}{rCl}
			\vect{J}_{\vect{\phi}} = \vect{J}_{\vect{\phi}}^{(\text{p})} + \vect{J}_{\vect{\phi}}^{(\text{o})}\label{eq:J_phi definition},
		\end{IEEEeqnarray}
		where $\vect{J}_{\vect{\phi}}^{(\text{p})}$ accounts for the prior information on $\vect{\phi}$ and $\vect{J}_{\vect{\phi}}^{(\text{o})}$ accounts for the observation-related FI. With $\epsilon_{\text{clk}}$ being the only parameter with prior information, it is straightforward to find that the only non-zero entry of $\vect{J}_{\vect{\phi}}^{(\text{p})}$ is 
		\begin{IEEEeqnarray}{rCl}
			\left[\vect{J}_{\vect{\phi}}^{(\text{p})}\right]_{1,1} &=& \E{-\frac{\partial^2 \ln p(\epsilon_{\text{clk}})}{\partial \epsilon_{\text{clk}}^2}}{ } = \frac{1}{\sigma_{\text{clk}}^2}, \label{eq:J_clk_prior}
		\end{IEEEeqnarray}
		where $p(\epsilon_{\text{clk}})$ is the probability density function (pdf) of $\epsilon_{\text{clk}}$.
		Since $\vect{\phi}$ is observed under Gaussian noise, the $(i,j)$-th entry of the $\vect{J}_{\vect{\phi}}^{(\text{o})}$ is
		\begin{IEEEeqnarray}{rCl}
			\left[\vect{J}_{\vect{\phi}}^{(\text{o})}\right]_{i,j} &=& \frac{1}{\sigma_{\eta,\text{R}}^2}\sum_{b=1}^{N_{\text{B}}} \sum_{p\in \mathcal{P}} \Re\left\{\frac{\partial \vect{m}_b\herm[p]}{\partial \phi_i} \left(\vect{F}_{\text{R}}\herm\vect{F}_{\text{R}}\right)^{-1} \frac{\partial \vect{m}_b[p]}{\partial \phi_j}\right\},
			\label{eq:entries of channel parameter FIM}
			\IEEEeqnarraynumspace
		\end{IEEEeqnarray}
		where
		\begin{IEEEeqnarray}{rCl}
			{\vect{m}}_b[p] &=& {g}[p] \e^{-\jj \omega_p\tau_{\text{s}}} \vect{F}_{\text{R}}\herm \vect{H}[p]\vect{F}_{\text{T}}^* {\vect{s}}_b[p].
			\IEEEeqnarraynumspace
		\end{IEEEeqnarray}
		The required derivatives of ${\vect{m}}_b[p]$ are provided in Appendix~\ref{sec:appendix - derivative of m_b[p] wrt to phi}.
		
		Instead of $\vect{\phi}$, which includes the unknown channel parameters, we are rather interested in the estimation of a parameter vector including the unknown position parameters. 
		We consider two cases, depending on the available knowledge on the Tx and Rx position and orientation. In the first case, the Tx position $\vect{p}_{\text{T}}$ and orientation $\alpha_{\text{T}}$ are known and the Rx position $\vect{p}_{\text{R}}$ and orientation $\alpha_{\text{R}}$ are unknown (receiver localization). The position parameter vector then reads as
		\begin{IEEEeqnarray}{rCl}
			\tilde{\vect{\phi}}_{\text{R}} &=& \begin{cases}
				[\vect{p}_{\text{R}}\trans, \alpha_{\text{R}}, \epsilon_{\text{clk}}, h_{0,\Re}, h_{0,\Im}, \ldots, \vect{p}_{\text{s},L-1}\trans, h_{L-1,\Re}, h_{L-1,\Im}
				]\trans\in\mathbb{R}^{4L + 2}, & \text{if LOS exists,}\\
				[\vect{p}_{\text{R}}\trans, \alpha_{\text{R}}, \epsilon_{\text{clk}}, \vect{p}_{\text{s},0}, h_{0,\Re}, h_{0,\Im}, \ldots, \vect{p}_{\text{s},L-1}\trans, h_{L-1,\Re}, h_{L-1,\Im}
				]\trans\in\mathbb{R}^{4L + 4}, & \text{otherwise}.
			\end{cases}\label{eq:position parameter vector LOS DL}
			\IEEEeqnarraynumspace
		\end{IEEEeqnarray}
		In the second case, the Rx estimates the Tx position and orientation, knowing its own position and orientation (transmitter localization). The parameter vector is then
		\begin{IEEEeqnarray}{rCl}
			\tilde{\vect{\phi}}_{\text{T}} &=& \begin{cases}
				[\vect{p}_{\text{T}}\trans, \alpha_{\text{T}}, \epsilon_{\text{clk}}, h_{0,\Re}, h_{0,\Im}, \ldots, \vect{p}_{\text{s},L-1}\trans, h_{L-1,\Re}, h_{L-1,\Im}
				]\trans\in\mathbb{R}^{4L + 2}, & \text{if LOS exists,}\\
				[\vect{p}_{\text{T}}\trans, \alpha_{\text{T}}, \epsilon_{\text{clk}}, \vect{p}_{\text{s},0}, h_{0,\Re}, h_{0,\Im}, \ldots, \vect{p}_{\text{s},L-1}\trans, h_{L-1,\Re}, h_{L-1,\Im}
				]\trans\in\mathbb{R}^{4L + 4}, & \text{otherwise}.
			\end{cases}\label{eq:position parameter vector LOS UL}
			\IEEEeqnarraynumspace
		\end{IEEEeqnarray}
		
		The FIMs $\vect{J}_{\tilde{\vect{\phi}}_{\text{R}}}, \vect{J}_{\tilde{\vect{\phi}}_{\text{T}}}$ can be obtained using $\vect{J}_{\vect{\phi}}$ and the transformation matrices $\vect{T}_{\text{R}}, \vect{T}_{\text{T}}$~\cite{LC98}: 
		\begin{IEEEeqnarray}{rCl}
			\vect{J}_{\tilde{\vect{\phi}}_{\text{R}}} &=& \vect{T}_{\text{R}}\vect{J}_{\vect{\phi}}\vect{T}_{\text{R}}\trans,\\
			\vect{J}_{\tilde{\vect{\phi}}_{\text{T}}} &=& \vect{T}_{\text{T}}\vect{J}_{\vect{\phi}}\vect{T}_{\text{T}}\trans,
			\IEEEeqnarraynumspace
		\end{IEEEeqnarray}
		where $	\left[\vect{T}_{\text{R}}\right]_{i,j} = \partial [\vect{\phi}]_j/\partial [\tilde{\vect{\phi}}_{\text{R}}]_i$ and $[\vect{T}_{\text{T}}]_{i,j} = \partial [\vect{\phi}]_j/\partial [\tilde{\vect{\phi}}_{\text{T}}]_i$. The entries of $\vect{T}_{\text{R}}, \vect{T}_{\text{T}}$ are given in Appendix~\ref{sec:appendix - entries of Fisher information transformation matrices}. In the following, we will refer to both $\vect{J}_{\tilde{\vect{\phi}}_{\text{R}}}$ and $\vect{J}_{\tilde{\vect{\phi}}_{\text{T}}}$ as $\vect{J}_{\tilde{\vect{\phi}}}$ and to $\vect{T}_{\text{R}}$ and $\vect{T}_{\text{T}}$ as $\vect{T}$, explicitly distinguishing between them where necessary.
		
		We use the notion of the equivalent FIM (EFIM)~\cite{SW07} to focus on the available information on the position and orientation parameters. The EFIM can have much lower dimensions than $\vect{J}_{\tilde{\vect{\phi}}}$, but still fully describes the available information on the parameters of interest. Splitting $\vect{T}$ as $\vect{T} = \big[\vect{T}_{\text{po}}\trans,\; \vect{T}_{\text{np}}\trans\big]\trans$, with $\vect{T}_{\text{po}}$ consisting of the first three rows of $\vect{T}$, which correspond to the position and orientation parameters, and $\vect{T}_{\text{np}}$ containing the rest, the EFIM for the position and orientation parameters is
		\begin{IEEEeqnarray}{rCl}
			\vect{J}_{\text{po}} &=& \vect{T}_{\text{po}} \vect{J}_{\vect{\phi}} \vect{T}_{\text{po}}\trans - \vect{T}_{\text{po}} \vect{J}_{\vect{\phi}} \vect{T}_{\text{np}}\trans \left(\vect{T}_{\text{np}} \vect{J}_{\vect{\phi}}\vect{T}_{\text{np}}\trans\right)^{-1}\vect{T}_{\text{np}} \vect{J}_{\vect{\phi}} \vect{T}_{\text{po}}\trans.\label{eq:J_po_definition}
			\IEEEeqnarraynumspace
		\end{IEEEeqnarray}
		The measure we use for characterizing the achievable position estimation accuracy is the position error bound (PEB) defined as~\cite{SGD+15} 
		\begin{IEEEeqnarray}{rCl}
			\text{PEB} = \sqrt{\left[\vect{J}_{\text{po}}^{-1}\right]_{1,1} + \left[\vect{J}_{\text{po}}^{-1}\right]_{2,2}}.\label{eq:PEB definition}
		\end{IEEEeqnarray}
		
		\subsection{Dynamic Scenario}
		In the dynamic scenario the channel parameter vector $\vect{\phi}$ is extended from \eqref{eq:channel parameter vector} to
		\begin{IEEEeqnarray}{rCl}
			\vect{\phi} &=& [\epsilon_{\text{clk}}, \tau_0, \tilde{\theta}_{\text{T},0}, \tilde{\theta}_{\text{R},0}, v_{0},h_{0, \Re}, h_{0, \Im}, \ldots, \tau_{L-1}, \tilde{\theta}_{\text{T},{L-1}}, \tilde{\theta}_{\text{R},{L-1}}, v_{L-1}, h_{{L-1},\Re}, h_{{L-1},\Im}
			]\trans\hspace*{-0.1cm} \in \mathbb{R}^{6L+1}.
			\IEEEeqnarraynumspace
		\end{IEEEeqnarray}
		For receiver localization, the position parameter vector is defined as
		\begin{IEEEeqnarray}{rCl}
			\tilde{\vect{\phi}}_{\text{R}} &=& \begin{cases}
				[\vect{p}_{\text{R}}\trans, \alpha_{\text{R}}, \vect{v}_{\text{R}}, \epsilon_{\text{clk}}, h_{0,\Re}, h_{0,\Im}, \ldots, \vect{p}_{\text{s},L-1}\trans, h_{L-1,\Re}, h_{L-1,\Im}
				]\trans\hspace*{-0.05cm}\in\mathbb{R}^{4L + 4}, & \hspace*{-0.45cm}\text{if LOS exists},\\
				[\vect{p}_{\text{R}}\trans, \alpha_{\text{R}}, \vect{v}_{\text{R}}, \epsilon_{\text{clk}}, \vect{p}_{\text{s},0}, h_{0,\Re}, h_{0,\Im}, \ldots, \vect{p}_{\text{s},L-1}\trans, h_{L-1,\Re}, h_{L-1,\Im}
				]\trans\hspace*{-0.05cm}\in\mathbb{R}^{4L + 6}, & \hspace*{-0.25cm}\text{otherwise}.
			\end{cases}\label{eq:position parameter vector dynamic}
			\IEEEeqnarraynumspace
		\end{IEEEeqnarray}
		For transmitter localization the position parameter vector can be defined accordingly. 
		We are interested in the EFIM $\vect{J}_{\text{pov}}$ of the position, orientation and velocity parameters:
		\begin{IEEEeqnarray}{rCl}
			\vect{J}_{\text{pov}} &=& \vect{T}_{\text{pov}} \vect{J}_{\vect{\phi}} \vect{T}_{\text{pov}}\trans - \vect{T}_{\text{pov}} \vect{J}_{\vect{\phi}} \vect{T}_{\text{np}}\trans \left(\vect{T}_{\text{np}} \vect{J}_{\vect{\phi}}\vect{T}_{\text{np}}\trans\right)^{-1}\vect{T}_{\text{np}} \vect{J}_{\vect{\phi}} \vect{T}_{\text{pov}}\trans,
			\IEEEeqnarraynumspace
		\end{IEEEeqnarray}
		where we have split $\vect{T}$ as $\vect{T} = \big[\vect{T}_{\text{pov}}\trans, \; \vect{T}_{\text{np}}\trans\big]\trans$, with $\vect{T}_{\text{pov}}$ consisting of the first 5 rows of $\vect{T}$.

	\section{Downlink vs Uplink Positioning}
		\label{sec:downlink and uplink positioning}
		In the following, we discuss the positioning and orientation estimation of a user equipment (UE) in a communication system, assuming the position and orientation of the base station (BS) are known. In constrast to \cite{AZA+18}, we assume
		that the devices (BS or UE) use the same set of beamforming vectors for transmission and reception, e.g. $\vect{F}_{\text{R}}$ in the DL is equal to $\vect{F}_{\text{T}}$ in the UL. We consider this setup because the presence of $\vect{F}_{\text{T}}$ and $\vect{F}_{\text{R}}$ usually results from some hardware constraints, which are device-specific and independent of whether the device is in Tx or Rx mode. For example, due to the limited number of RF chains, a UE might only be able to process a signal in Tx and Rx mode with a set of analog beamforming vectors. We consider here the static scenario, but a similar analysis can be conducted in the dynamic scenario. 
		With known BS position and orientation, positioning of a UE in the DL corresponds to receiver localization, whereas positioning of a UE in the UL corresponds to transmitter localization.
		Thus, by replacing the subscripts 'T' and 'R' with 'BS' and 'UE' in \eqref{eq:signal model} and \eqref{eq:static channel definition} for the DL and vice-versa for the UL, we obtain the signal model for the DL and the UL, respectively:
		\begin{IEEEeqnarray}{rrCl}
			\text{(DL)}:\quad&\vect{w}_b^{(\text{DL})}[p] &\approx& \vect{F}_{\text{UE}}\herm \vect{H}^{(\text{DL})}[p] \vect{F}_{\text{BS}}^* \vect{s}_{b}^{(\text{DL})}[p] + \vect{F}_{\text{UE}}\herm\vect{\eta}_b[p]\label{eq:signal model DL}\\
			\text{(UL)}:\quad&\vect{w}_b^{(\text{UL})}[p] &\approx& \vect{F}_{\text{BS}}\herm \vect{H}^{(\text{UL})}[p] \vect{F}_{\text{UE}}^* \vect{s}_{b}^{(\text{UL})}[p] + \vect{F}_{\text{BS}}\herm\vect{\eta}_b[p]\label{eq:signal model UL},
		\end{IEEEeqnarray}
		where we have that  $\vect{H}^{(\text{UL})}[p] = \left(\vect{H}^{(\text{DL})}[p]\right)\trans$ and that the BS and UE use the same beamforming vectors for transmission and reception, as mentioned above. From the parameter vector for receiver localization \eqref{eq:position parameter vector LOS DL} and the corresponding parameter vector for transmitter localization \eqref{eq:position parameter vector LOS UL}, it is straightforward to verify that $\tilde{\vect{\phi}}^{(\text{DL})} = \tilde{\vect{\phi}}^{(\text{UL})}$.
		
		We now compare the positioning accuracy between the DL and the UL, under the following assumptions:
		\begin{enumerate}
			\item $\vect{F}_{\text{UE}}$ and $\vect{F}_{\text{BS}}$ have orthonormal columns.
			\item The signals transmitted through different beamforming vectors are uncorrelated and have equal power, i.e. $\E{\vect{s}_b^{(\text{DL})}[p]\big(\vect{s}_b^{(\text{DL})}[p]\big)\herm}{} = \gamma_p P_{\text{BS}}/M_{\text{BS}}$ and $\E{\vect{s}_b^{(\text{UL})}[p]\big(\vect{s}_b^{(\text{UL})}[p]\big)\herm}{} = \gamma_p P_{\text{UE}}/M_{\text{UE}}$, where $\gamma_p$ is the fraction of the power of the OFDM symbol that is spent on the $p$-th subcarrier, with $\sum_{p\in\mathcal{P}}\gamma_p=1$. 
		\end{enumerate}
		$\vect{J}_{\tilde{\vect{\phi}}}$ in the DL and the UL can be obtained in a similar fashion as $\vect{J}_{\vect{\phi}}$ in \eqref{eq:J_phi definition}, \eqref{eq:J_clk_prior} and \eqref{eq:entries of channel parameter FIM}. We define $\vect{Z}$ as the matrix which has the same dimensions as $\vect{J}_{\tilde{\vect{\phi}}}$ and its $(i,j)$-th element is
		\begin{IEEEeqnarray}{rCl}
			\left[\vect{Z}\right]_{i,j} &=& \trace\left(\Re\left\{\sum_{b=1}^{N_{\text{B}}} \sum_{p\in\mathcal{P}}\vect{F}_{\text{BS}}\trans \frac{\partial \left(\vect{H}^{(\text{DL})}[p]\right)\herm}{\partial \tilde{\phi}_i} \vect{F}_{\text{UE}} \vect{F}_{\text{UE}}\herm \frac{\partial \vect{H}^{(\text{DL})}[p]}{\partial \tilde{\phi}_j} \vect{F}_{\text{BS}}^* \right\}\right).
		\end{IEEEeqnarray}
		From \eqref{eq:J_phi definition}, \eqref{eq:J_clk_prior}, \eqref{eq:entries of channel parameter FIM}, \eqref{eq:signal model DL} and taking the expectation over $\vect{s}_b^{(\text{DL})}[p]$ we can compute $\vect{J}_{\tilde{\vect{\phi}}}^{(\text{DL})}$ as
		\begin{IEEEeqnarray}{rCl}
			\vect{J}_{\tilde{\vect{\phi}}}^{(\text{DL})} &=& \frac{P_{\text{BS}}}{\sigma_{\eta,\text{UE}}^2 M_{\text{BS}}}\vect{Z} + \vect{e}_4\vect{e}_4\trans\frac{1}{\sigma_{\text{clk}}^2}\label{eq:J_phi_til_DL},
		\end{IEEEeqnarray}
		where $\sigma_{\eta,\text{UE}}^2$ is the noise variance at the UE in the DL and $\vect{e}_i$ is the vector whose $i$-th entry is equal to $1$ and the rest of its entries are $0$.
		Given that $\vect{H}_{b,p}^{(\text{UL})} = \big(\vect{H}_{b,p}^{(\text{DL})}\big)\trans$ we can show by using simple properties of the trace operator that
		\begin{IEEEeqnarray}{rCl}
			\vect{J}_{\tilde{\vect{\phi}}}^{(\text{UL})} &=&\frac{P_{\text{UE}}}{\sigma_{\eta,\text{BS}}^2 M_{\text{UE}}} \vect{Z} +  \vect{e}_4\vect{e}_4\trans\frac{1}{\sigma_{\text{clk}}^2},\label{eq:J_phi_til_UL}
		\end{IEEEeqnarray}
		where $\sigma_{\eta,\text{UE}}^2$ is the noise variance at the BS in the UL. Using the Sherman-Morrison formula we invert $\vect{J}_{\tilde{\vect{\phi}}}^{(\text{DL})}$ and $\vect{J}_{\tilde{\vect{\phi}}}^{(\text{UL})}$ and after a few steps we find that the ratio of the DL and UL PEB (see \eqref{eq:PEB definition}) is given by
		\begin{IEEEeqnarray}{rCl}
			\frac{\text{PEB}^{(\text{DL})}}{\text{PEB}^{(\text{UL})}} &=& \sqrt{\frac{\delta_{\text{UL}}}{\delta_{\text{DL}}} \frac{\zeta  - \frac{1}{\delta_{\text{DL}}\sigma_{\text{clk}}^2 + [\vect{Z}^{-1}]_{4,4}} }{\zeta  - \frac{1}{\delta_{\text{UL}}\sigma_{\text{clk}}^2 + [\vect{Z}^{-1}]_{4,4}} }}, \label{eq:DL and UL PEB ratio with sync error}
		\end{IEEEeqnarray}
		where $\zeta = \left([\vect{Z}^{-1}]_{1,1} + [\vect{Z}^{-1}]_{2,2}\right) /\big(\left([\vect{Z}^{-1}]_{1,4}\right)^2 + \left([\vect{Z}^{-1}]_{2,4}\right)^{2}\big)$ and $\delta_{\text{DL}} = P_{\text{BS}}M_{\text{UE}}/\sigma_{\eta,\text{UE}}^2$ and $\delta_{\text{UL}} = P_{\text{UE}}M_{\text{BS}}/\sigma_{\eta,\text{BS}}^2$ are the DL and UL receive SNR, respectively. It follows from \eqref{eq:DL and UL PEB ratio with sync error} that the link with the highest receive SNR will provide the best positioning performance. For perfectly synchronized ($\sigma_{\text{clk}}^2=0$) or asynchronous ($\sigma_{\text{clk}}^2\rightarrow\infty$) clocks we have
		\begin{IEEEeqnarray}{rCl}
			\frac{\text{PEB}^{(\text{DL})}}{\text{PEB}^{(\text{UL})}} &=& \sqrt{\frac{\delta_{\text{UL}}}{\delta_{\text{DL}}}}.\label{eq:DL and UL PEB ratio without sync error}
		\end{IEEEeqnarray}
		From \eqref{eq:DL and UL PEB ratio with sync error} and \eqref{eq:DL and UL PEB ratio without sync error} we see that DL and UL positioning have different accuracy, owing to different receive SNR. When $\delta_{\text{DL}} = \delta_{\text{UL}}$, again from \eqref{eq:DL and UL PEB ratio with sync error}, we find that $\text{PEB}^{(\text{DL})} = \text{PEB}^{(\text{UL})}$. We stress that \eqref{eq:DL and UL PEB ratio with sync error} and \eqref{eq:DL and UL PEB ratio without sync error} hold when each of the devices (BS or UE) employs the same set of orthonormal beamforming vectors for transmission and reception.
		In case the sets of beamforming vectors for transmission and reception are not the same, there is a mismatch between the subspaces spanned for transmission and reception. Although an analytic result has not been derived for this case, in Section~\ref{sec:numerical results}, we compare DL and UL positioning using random beamforming matrices for transmission and reception.
		
	\section{Asymptotic Analysis and Geometric Interpretation of Fisher Information}
		\label{sec:geometric interpretation of Fisher information}
		The expressions for the FIMs and their entries provided in Section~\ref{sec:CRLB computation} and in Appendix~\ref{sec:appendix - entries of Fisher information transformation matrices}, although useful for numerical evaluation of the achievable accuracy, do not provide much intuition about the positioning information that is available in the considered single-anchor MIMO-OFDM setup. Our aim is to obtain a geometric interpretation of the Fisher information, as in~\cite{HSZ+16, MWB+17}. Following~\cite{AZA+18,SGD+18,MWB+17}, we will consider asymptotic expressions for the position and orientation EFIM for large bandwidth\footnote{In an OFDM system with a given number of subcarriers and a fixed set of occupied subcarriers, the occupied bandwidth increases linearly with the sampling rate. Hence, equivalently to writing that the bandwidth goes to infinity we write $F_{\text{s}}\rightarrow \infty$} ($F_{\text{s}}\rightarrow \infty$) and large number of transmit and receive antennas ($N_{\text{T}}, N_{\text{R}} \rightarrow \infty$). The accuracy of the asymptotic expressions has been investigated in \cite{AZA+18}, where it was shown that they can provide a very accurate approximation of the exact PEB, even for moderate bandwidth and number of transmit and receive antennas. 
		
		Since, for arrays with a fixed aperture the asymptotic orthogonality/favorable propagation condition~\cite{Che13, NLM14} may not be satisfied~\cite{MM15}, we consider arrays whose aperture grows with increasing number of elements.In addition to the standard favorable propagation condition
		\begin{IEEEeqnarray}{rCl}
			\lim_{N_{\text{T}}\rightarrow\infty}\vect{a}_{\text{T},p}\herm(\tilde{\theta}_{\text{T},l})\vect{a}_{\text{T},p}(\tilde{\theta}_{\text{T},l'})/N_{\text{T}} = 0, \label{eq:fav_prop asymptotic orthogonality}
		\end{IEEEeqnarray}
		we also assume 
		\begin{IEEEeqnarray}{rCl}
			\lim_{N_{\text{T}}\rightarrow\infty}\frac{\partial}{\partial\tilde{\theta}_{\text{T},l}} \vect{a}_{\text{T},p}\herm(\tilde{\theta}_{\text{T},l})\vect{a}_{\text{T},p}(\tilde{\theta}_{\text{T},l'})/N_{\text{T}}^2 = 0,\label{eq:fav_prop first derivative}\\ \lim_{N_{\text{T}}\rightarrow\infty}\frac{1}{N_{\text{T}}^3}\big\|\partial\vect{a}_{\text{T},p}(\tilde{\theta}_{\text{T},l})/\partial\tilde{\theta}_{\text{T},l}  \big\|_2^2 > 0\label{eq:fav_prop second derivative},
		\end{IEEEeqnarray}
		implying that asymptotic orthogonality holds also for the inner product of the derivative of the array steering vector and the array steering vector at any angle. Similar assumptions are made for the Rx array. We can verify that these conditions hold both for ULAs and UCAs.
		
		In order to focus on the characteristics of the channel and obtain geometrically intuitive expressions for the Fisher information, we also use the following simplifying assumptions: 
		\begin{enumerate}
			\item $\tilde{g}(t)$ has a flat frequency response over the occupied subcarriers.
			\item $\vect{F}_{\text{T}} = \vect{I}_{N_{\text{T}}}$ and $\vect{F}_{\text{R}} = \vect{I}_{N_{\text{R}}}$, i.e. the whole transmit and receive spaces  can be sensed.
			\item The entries of the reference signal $\vect{x}_b[p]$ are assumed to be independent and identically distributed (i.i.d) random variables with $\E{\vect{x}_b[p] \vect{x}\herm_b[p]}{} = \gamma_p P_{\text{T}}/N_{\text{T}} \vect{I}_{N_{\text{T}}}$ and $\E{\vect{x}_b[p] \vect{x}\herm_b[q]}{} = \vect{0},\;p\neq q$. Hence, in this section we compute the entries of the FIMs, taking the expectation over the reference signal, as in~\cite{MWB+17}:
			\begin{IEEEeqnarray}{rCl}
				\left[\vect{J}_{\vect{\phi}}^{(\text{o})}\right]_{i,j} &=& \frac{1}{\sigma_{\eta,\text{R}}^2} \sum_{b=1}^{N_{\text{B}}} \sum_{p\in \mathcal{P}} \Re\left\{\E{\frac{\partial \vect{m}_b\herm[p]}{\partial \phi_i} \frac{\partial \vect{m}_b[p]}{\partial \phi_j}}{}\right\}.
				\label{eq:expectation of entries of channel parameter FIM}
				\IEEEeqnarraynumspace
			\end{IEEEeqnarray}
		\end{enumerate}
		To make the following expressions more compact we set $d_{\text{T},\text{R}} = \left\|\vect{p}_{\text{R}} - \vect{p}_{\text{T}}\right\|_2$,
		$d_{\text{T},\text{s},l} = \left\|\vect{p}_{\text{s},l} - \vect{p}_{\text{T}}\right\|_2$ and $d_{\text{R},\text{s},l} = \left\|\vect{p}_{\text{R}} - \vect{p}_{\text{s},l}\right\|_2$.
		As before, we distinguish between the static and dynamic scenario. 
		
		\subsection{Static Scenario}
		Before presenting the main results we define a few useful quantities.
		\begin{definition}
			The \emph{effective baseband bandwidth} $\beta$ of the signal is defined as
			\begin{IEEEeqnarray}{rCl}
				\beta &=& \sqrt{\sum\nolimits_{p\in\mathcal{P}}\gamma_p \omega_p^2 - \left(\sum\nolimits_{p\in\mathcal{P}}\gamma_p \omega_p\right)^2 }.
				\label{eq:effective baseband bandwidth}\IEEEeqnarraynumspace
			\end{IEEEeqnarray}
		\end{definition}
		\begin{definition}
			The \emph{effective angular carrier frequency} $\bar{\omega}_c$ of the signal is defined as
			\begin{IEEEeqnarray}{rCl}
				\bar{\omega}_c &=& \sqrt{\sum\nolimits_{p\in\mathcal{P}} \gamma_p \left(\omega_c + \omega_p\right)^2}.\label{eq:effective angular carrier frequency}\
				\IEEEeqnarraynumspace
			\end{IEEEeqnarray}
		\end{definition}
		The definitions in~\eqref{eq:effective baseband bandwidth} and~\eqref{eq:effective angular carrier frequency}
		are the multi-carrier counterparts of the effective baseband bandwidth and the effective carrier frequency defined in~\cite{HSZ+16} for single-carrier systems.
		\begin{definition}
			The \emph{squared array aperture function} (SAAF) of the Tx array is defined as~\cite{HSZ+16}
			\begin{IEEEeqnarray}{rCl}
				S_{\text{T}}(\tilde{\theta}_{\text{T},l}) &=& \frac{1}{N_{\text{T}}} \sum_{j=1}^{N_{\text{T}}}\left(d_{\text{T},j}\vect{u}_{\perp}\trans\left(\psi_{\text{T},j}\right)\vect{u}(\tilde{\theta}_{\text{T},l})\right)^2,
				\IEEEeqnarraynumspace
			\end{IEEEeqnarray}
			where $\vect{u}_{\perp}\left(\psi\right) = \vect{u}\left(\psi - \pi/2\right)$, with the array's centroid chosen at its reference point. The SAAF of the Rx array is defined accordingly. The Tx (Rx) SAAF fully describes the effect of the Tx (Rx) array structure on the AOD (AOA) information.
		\end{definition}
		
		\begin{theorem}
			\label{theor:receiver positioning without time stamps}
			\emph{(Static receiver localization)} The EFIM for the position $\vect{p}_{\text{R}}$ and orientation $\alpha_{\text{R}}$ of an imperfectly synchronized static receiver, when the position $\vect{p}_{\text{T}}$ and orientation $\alpha_{\text{T}}$ of the transmitter are known, can be asymptotically expressed as
			\begin{IEEEeqnarray}{rCl}
				\vect{J}_{\text{po}}^{(\text{s})} &\rightarrow& \begin{cases}
					\delta_{\text{R},\text{T}} \left(\vect{J}_{\text{LOS}}^{(\text{s})} + \sum_{l=1}^{L-1} \vect{J}_{\text{NLOS}, l}^{(\text{s})} - \vect{J}_{\epsilon_{\text{clk}}}^{(\text{s})}\right), & \text{if LOS exists,}\\
					\delta_{\text{R},\text{T}} \left(\sum_{l=0}^{L-1} \vect{J}_{\text{NLOS}, l}^{(\text{s})} - \vect{J}_{\epsilon_{\text{clk}}}^{(\text{s})}\right),& \text{otherwise},
				\end{cases}\label{eq:J_p receiver without time stamps}
				\IEEEeqnarraynumspace
			\end{IEEEeqnarray}
			where $\delta_{\text{R},\text{T}} = N_{\text{R}} N_{\text{B}} P_{\text{T}}/\sigma_{\eta,\text{R}}^2$ is the receive SNR. The information from the LOS path is given by
			\begin{IEEEeqnarray}{rCl}
				\vect{J}_{\text{LOS}}^{(\text{s})} &=&  \left|h_0\right|^2 \left[\frac{\beta^2}{c^2}\vect{z}_{\tau_0} \vect{z}_{\tau_0}\trans + \frac{\bar{\omega}_c^2}{c^2d_{\text{T},\text{R}}^2} \left(S_{\text{T}}(\tilde{\theta}_{\text{T},0}) \vect{z}_{\theta_{\text{T},0}} \vect{z}_{\theta_{\text{T},0}}\trans + S_{\text{R}}(\tilde{\theta}_{\text{R},0}) \vect{z}_{\theta_{\text{R},0}} \vect{z}_{\theta_{\text{R},0}}\trans \right)\right],
				\label{eq:J_LOS receiver no time-stamp}
				\IEEEeqnarraynumspace
			\end{IEEEeqnarray}
			with 
			$\vect{z}_{\tau_0} = [
			-\vect{u}\trans\left(\theta_{\text{R},0}\right), \; 0]\trans$,
			$\vect{z}_{\theta_{\text{T},0}} = [\vect{u}_{\perp}\trans\left(\theta_{\text{R},0}\right), \; 0]\trans$
			and 
			$\vect{z}_{\theta_{\text{R},0}} = [\vect{u}_{\perp}\trans\left(\theta_{\text{R},0}\right), \; -d_{\text{T},\text{R}}]\trans$.
			The information from the $l$-th NLOS path is described by
			\begin{IEEEeqnarray}{rCl}
				\vect{J}_{\text{NLOS},l}^{(\text{s})} &=& \left|h_l\right|^2 f_l \vect{z}_l\vect{z}_l\trans, \label{eq:J_NLOS receiver no time-stamp}
				\IEEEeqnarraynumspace
			\end{IEEEeqnarray}
			with 
			\begin{IEEEeqnarray}{rCl}
				f_l &=& \left(\frac{c^2\left(1 + \cos\left(\Delta\theta_l\right)\right)^2}{\bar{\omega}_c^2} \left( \frac{d_{\text{T},\text{s},l}^2}{S_{\text{T}}(\tilde{\theta}_{\text{T},l})}  + \frac{d_{\text{R},\text{s},l}^2}{S_{\text{R}}(\tilde{\theta}_{\text{R},l})} \right) + \frac{\sin^2\left(\Delta\theta_l\right)c^2}{\beta^2}\right)^{-1},\label{eq:J_NLOS intensity of information}\\
				\vect{z}_l &=& \begin{bmatrix}
					\vect{u}_{\perp}\trans\left(\theta_{\text{T},l}\right) + \vect{u}_{\perp}\trans\left(\theta_{\text{R},l}\right), & -\left(1 + \cos\left(\Delta\theta_l\right)\right)d_{\text{R},\text{s},l}
				\end{bmatrix}\trans\in\mathbb{R}^3,\label{eq:J_NLOS direction of information}
				\IEEEeqnarraynumspace
			\end{IEEEeqnarray}
			and the information loss due to the synchronization error, is described by
			\begin{IEEEeqnarray}{rCl}
				\vect{J}_{\epsilon_{\text{clk}}}^{(\text{s})} &=& \vect{z}_{\epsilon_{\text{clk}}} \vect{z}_{\epsilon_{\text{clk}}}\trans/K_{\epsilon_{\text{clk}}}, \label{eq:J_coupl receiver no time-stamp}
				\IEEEeqnarraynumspace\label{eq:J_p receiver info loss matrix}
			\end{IEEEeqnarray}
			where
			\begin{IEEEeqnarray}{rCl}
				\vect{z}_{\epsilon_{\text{clk}}} &=& \begin{cases}
					|h_0|^2\frac{\beta^2}{c^2}\vect{z}_{\tau_0} + \sum_{l=1}^L \left|h_l\right|^2 \sin\big(\Delta\theta_l\big) f_l \vect{z}_l, & \text{if LOS exists,}\\
					\sum_{l=0}^L \left|h_l\right|^2 \sin\big(\Delta\theta_l\big) f_l \vect{z}_l, & \text{otherwise,}
				\end{cases}\label{eq:J_p receiver info loss direction}\\
				K_{\epsilon_{\text{clk}}} &=& \begin{cases}
					\frac{1}{\delta_{\text{R},\text{T}}\sigma_{\text{clk}}^2} + \left|h_0\right|^2 \frac{\beta^2}{c^2} + \sum_{l'=1}^{L-1} \left|h_{l'}\right|^2 \sin^2\big(\Delta\theta_{l'}\big) f_{l'}, & \text{if LOS exists,}\\
					\frac{1}{\delta_{\text{R},\text{T}}\sigma_{\text{clk}}^2} + \sum_{l'=0}^{L-1} \left|h_{l'}\right|^2 \sin^2\big(\Delta\theta_{l'}\big) f_{l'}, & \text{otherwise}.
				\end{cases}\label{eq:J_p receiver info loss intenisty}
			\end{IEEEeqnarray}
		\end{theorem}
		\begin{IEEEproof}
			See Appendix~\ref{sec:proof of theorem for receiver positioning without time-stamps}.
		\end{IEEEproof}
		Some interesting remarks are to be made based on Theorem \ref{theor:receiver positioning without time stamps}.
		Similar to~\cite{MWB+17}, we find that the EFIM of each of the single-bounce NLOS paths is rank-$1$. We can also see from \eqref{eq:J_NLOS receiver no time-stamp} and \eqref{eq:J_NLOS direction of information} that the direction of the position information offered by a single-bounce NLOS path is determined solely by the geometry and is independent of the system parameters. The direction is given by the sum of the 2D unit vectors pointing to the directions orthogonal to the AOD and AOA of the NLOS path.
		
		We observe that, in the LOS-only case, where the derived expressions \eqref{eq:J_p receiver without time stamps},\eqref{eq:J_LOS receiver no time-stamp}, \eqref{eq:J_coupl receiver no time-stamp}-\eqref{eq:J_p receiver info loss intenisty} are exact, for an imperfectly synchronized system ($0 < \sigma_{\text{clk}}^2<\infty$) positioning is possible. However, the ranging information provided by the TOA of the LOS path decreases with increasing $\sigma_{\text{clk}}$ and the achievable positioning accuracy is lower bounded by the synchronization error-induced ranging error $\sigma_{\text{clk}} c$. Hence, large values of $\sigma_{\text{clk}}$ may lead to poor position estimation accuracy. For an asynchronous system ($\sigma_{\text{clk}}^2\rightarrow\infty$), the ranging information from the TOA of the LOS path vanishes entirely and at least one NLOS path in addition to the LOS path is required.
		
		When there is no LOS path, at least three NLOS paths are required for positioning~\cite{MWB+17}. Similarly to the LOS-only case, the achievable accuracy is lower bounded by $\sigma_{\text{clk}}c$ and for an asynchronous system an additional NLOS path is required for positioning.
		
		Similar results can be obtained for unknown Tx position and orientation:
		\begin{theorem}
			\label{theor:transmitter positioning without time stamps}
			\emph{(Static transmitter localization)} The EFIM for the position $\vect{p}_{\text{T}}$ and the orientation $\alpha_{\text{T}}$ of an imperfectly synchronized transmitter, when the position $\vect{p}_{\text{R}}$ and orientation $\alpha_{\text{R}}$ of the receiver are known, can be asymptotically expressed as in \eqref{eq:J_p receiver without time stamps}, \eqref{eq:J_LOS receiver no time-stamp}, \eqref{eq:J_NLOS receiver no time-stamp} and \eqref{eq:J_coupl receiver no time-stamp}, with 
			$\vect{z}_{\tau_0} = [-\vect{u}\trans\left(\theta_{\text{T},0}\right), \; 0]\trans\in\mathbb{R}^3$, 
			$\vect{z}_{\theta_{\text{T},0}} = [\vect{u}_{\perp}\trans\left(\theta_{\text{T},0}\right), \; -d_{\text{T},\text{R}}]\trans$, 
			$\vect{z}_{\theta_{\text{R},0}} = [\vect{u}_{\perp}\trans\hspace*{-0.03cm}\left(\theta_{\text{T},0}\right), \; 0]\trans$, 
			and $\vect{z}_l = [\vect{u}_{\perp}\trans\hspace*{-0.03cm}\left(\theta_{\text{T},l}\right)\hspace*{-0.03cm} + \hspace*{-0.03cm}\vect{u}_{\perp}\trans\hspace*{-0.03cm}\left(\theta_{\text{R},l}\right), \; -\left(1\hspace*{-0.02cm} + \hspace*{-0.02cm}\cos\left(\Delta\theta_l\right)\right)d_{\text{T},\text{s},l}
			]\trans.$
		\end{theorem}
		\begin{IEEEproof}
			Similar to the proof of Theorem~\ref{theor:receiver positioning without time stamps}.
		\end{IEEEproof}
		Theorems~\ref{theor:receiver positioning without time stamps} and \ref{theor:transmitter positioning without time stamps} can be used for the analysis of the effect of the geometry and the system parameters on the DL and UL positioning accuracy, respectively. Theorem~\ref{theor:transmitter positioning without time stamps} can also be used for the analysis of sidelink positioning, where a UE estimates another UE's relative position.
		
		\subsection{Dynamic Scenario}
		In order to keep the expressions relatively simple and gain insight on the effect of the movement of the receiver or the transmitter, apart from the assumptions made for the static scenario, we additionally assume that $\omega_c \gg \omega_p, \; p\in\mathcal{P}$, in other words that the system under consideration is narrowband.
		Again, before presenting the results, we define a couple of useful quantities.
		\begin{definition}
			The \emph{effective baseband bandwidth} of the signal arriving from the $l$-th path is 
			\begin{IEEEeqnarray}{rCl}
				\beta_l &=& \sqrt{\sum\nolimits_{q\in\mathcal{P}}\xi_{l,q}\gamma_q \omega_q^2/\xi_{l} - \left(\sum\nolimits_{q\in\mathcal{P}}\xi_{l,q}\gamma_q \omega_p/\xi_{l}\right)^2},
				\label{eq:effective baseband bandwidth dynamic}
				\IEEEeqnarraynumspace
			\end{IEEEeqnarray}
			where $\xi_{l,q} = \sum\nolimits_{p\in\mathcal{P}} \left|Q\left(\Phi_{p,q,l}\right)\right|^2$ describes the intensity effect of the Doppler shift for the $l$-th path on the $q$-th subcarrier and $\xi_{l} = \sum\nolimits_{q\in\mathcal{P}}\gamma_q\xi_{l,q}$, with $\Phi_{p,q,l} = \left(\omega_p - \omega_q - v_{l}\omega_c/c\right)T_{\text{s}}/2$. Similar to what is reported in~\cite{HSZ+16}, the intensity effect for mm-Wave systems with large bandwidth is very small and can be neglected, i.e. $\xi_{l,q} \approx 1, \;l,q$. Then, \eqref{eq:effective baseband bandwidth dynamic} falls back to \eqref{eq:effective baseband bandwidth} $\forall l$.
		\end{definition} 
		\begin{definition}
			The \emph{root mean square (rms) duration} of the reference signal arriving from the $l$-th path is
			\begin{IEEEeqnarray}{rCl}
				t_{\text{rms},l} = T_{\text{s}}\sqrt{M^2\left(N_{\text{B}}^2 - 1\right)/12 + n_{\text{B},\text{rms},l}^2},
				\label{eq:t_rms}
			\end{IEEEeqnarray}
			where
			\begin{IEEEeqnarray}{rCl}
				n_{\text{B},\text{rms},l}^2 &=& \frac{\sum_{q\in\mathcal{P}}\gamma_q \sum\nolimits_{p\in\mathcal{P}} \left|Q\left(\Phi_{p,q,l}\right)\right|^2 \left(\cot\left(\Phi_{p,q,l}\right) - N\cot\left(N\Phi_{p,q,l}\right)\right)^2/4}{\xi_{l}}\nonumber\\
				&& - \left( \frac{\sum_{q\in\mathcal{P}}\gamma_q \sum\nolimits_{p\in\mathcal{P}} \left|Q\left(\Phi_{p,q,l}\right)\right|^2 \left(\cot\left(\Phi_{p,q,l}\right) - N\cot\left(N\Phi_{p,q,l}\right)\right)/2}{\xi_{l}} \right)^2.
			\end{IEEEeqnarray}
			It can be shown that for practical scenarios with $\|\vect{v}_{\text{R}}\|_2/c\ll 1$
			\begin{IEEEeqnarray}{rCl}
				n_{\text{B},\text{rms},l}^2 &\approx& n_{\text{B},\text{rms}}^2 = \sum_{q\in\mathcal{P}}\gamma_q\sum_{\substack{p\in\mathcal{P},\;p\neq q}}\sin^{-2}\left(\pi\left(p-q\right)/N\right).
			\end{IEEEeqnarray}
		\end{definition}
		We now present the main result for the dynamic scenario:
		\begin{theorem}
			\label{theor:receiver positioning without time stamps dynamic}
			\emph{(Mobile receiver localization)} The EFIM for the position $\vect{p}_{\text{R}}$, orientation $\alpha_{\text{R}}$ and velocity $\vect{v}_{\text{R}}$ of an imperfectly synchronized mobile receiver, when the position $\vect{p}_{\text{T}}$ and orientation $\alpha_{\text{T}}$ of the static transmitter are known, can be asymptotically expressed as
			\begin{IEEEeqnarray}{rCl}
				\vect{J}_{\text{pov}}^{(\text{d})} &\rightarrow& \begin{cases}
					\delta_{\text{R},\text{T}} \left(\vect{J}_{\text{LOS}}^{(\text{d})} + \sum_{l=1}^{L-1} \vect{J}_{\text{NLOS}, l}^{(\text{d})} - \vect{J}_{\epsilon_{\text{clk}}}^{(\text{d})}\right), & \text{if LOS exists,}\\
					\delta_{\text{R},\text{T}} \left(\sum_{0=1}^{L-1} \vect{J}_{\text{NLOS}, l}^{(\text{d})} - \vect{J}_{\epsilon_{\text{clk}}}^{(\text{d})}\right), & \text{otherwise},
				\end{cases} \label{eq:J_p receiver without time stamps dynamic}
				\IEEEeqnarraynumspace
			\end{IEEEeqnarray}
			where the information from the LOS path is described by
			\begin{IEEEeqnarray}{rCl}
				\vect{J}_{\text{LOS}}^{(\text{d})} \hspace*{-0.03cm}&=&\hspace*{-0.03cm}  \left|h_0\right|^2 \hspace*{-0.02cm}\xi_0 \hspace*{-0.03cm}\left[ \frac{\beta_0^2}{c^2} \vect{z}_{\tau_0} \vect{z}_{\tau_0}\trans\hspace*{-0.02cm} + \hspace*{-0.02cm}\frac{\omega_c^2}{c^2d_{\text{T},\text{R}}^2}\hspace*{-0.02cm} \left(\hspace*{-0.02cm}S_{\text{T}}\hspace*{-0.01cm}(\hspace*{-0.01cm}\tilde{\theta}_{\text{T},0}\hspace*{-0.01cm})\hspace*{-0.02cm} \vect{z}_{\theta_{\text{T},0}} \vect{z}_{\theta_{\text{T},0}}\trans \hspace*{-0.02cm}+\hspace*{-0.02cm} S_{\text{R}\hspace*{-0.01cm}}(\hspace*{-0.01cm}\tilde{\theta}_{\text{R},0}\hspace*{-0.01cm})\hspace*{-0.01cm} \vect{z}_{\theta_{\text{R},0}} \vect{z}_{\theta_{\text{R},0}}\trans \hspace*{-0.02cm}+\hspace*{-0.02cm} \rho_{0}^2 t_{\text{rms},0}^2\vect{z}_{v_0} \vect{z}_{v_0}\trans\hspace*{-0.02cm}\right)\hspace*{-0.03cm}\right]\hspace*{-0.04cm},
				\label{eq:J_LOS receiver no time-stamp dynamic}
				\IEEEeqnarraynumspace
			\end{IEEEeqnarray}
			with
			\begin{IEEEeqnarray}{rCl}
				\vect{z}_{\tau_0} &=& [-\vect{u}\trans\left(\theta_{\text{R},0}\right), \; 0, \; 0, \; 0]\trans\in\mathbb{R}^5,\label{eq:v_LOS,tau receiver dynamic}\\
				\vect{z}_{\theta_{\text{T},0}} &=& [\vect{u}_{\perp}\trans\left(\theta_{\text{R},0}\right), \; 0, \; 0, \; 0]\trans\in\mathbb{R}^5,\label{eq:v_LOS,T receiver dynamic}\\
				\vect{z}_{\theta_{\text{R},0}} &=& [\vect{u}_{\perp}\trans\left(\theta_{\text{R},0}\right), \; -d_{\text{T},\text{R}}, \; 0, \; 0]\trans\in\mathbb{R}^5,\label{eq:v_LOS,R receiver dynamic}\\
				\vect{z}_{v_0} &=& [-\vect{u}_{\perp}\trans\left(\theta_{\text{R},0}\right), \; 0, \; d_{\text{T},\text{R}} \vect{u}\trans\left(\theta_{\text{R},0}\right)/\rho_{0}]\trans\in\mathbb{R}^5\label{eq:v_LOS,v receiver dynamic},
				\IEEEeqnarraynumspace
			\end{IEEEeqnarray}
			and $\rho_l = \vect{v}_{\text{R}}\trans \vect{u}_{\perp}\left(\theta_{\text{R},l}\right), \;l=0,\ldots,L-1,$ being the transverse velocity component.
			The information from the $l$-th NLOS path neglecting its coupling with other paths is described by
			\begin{IEEEeqnarray}{rCl}
				\vect{J}_{\text{NLOS},l}^{(\text{d})} &=& \left|h_l\right|^2 \xi_l\left( f_{\tau, \theta, l} \vect{z}_{\tau, \theta, l}\vect{z}_{\tau, \theta, l}\trans + f_{\tau, v, l} \vect{z}_{\tau, v, l}\vect{z}_{\tau, v, l}\trans + f_{\theta, v, l} \vect{z}_{\theta, v, l}\vect{z}_{\theta, v, l}\trans \right), \label{eq:J_NLOS receiver no time-stamp dynamic}
				\IEEEeqnarraynumspace
			\end{IEEEeqnarray}
			with 
			\begin{IEEEeqnarray}{rCl}
				\vect{z}_{\tau, \theta, l} &=& [\vect{u}_{\perp}\trans\left(\theta_{\text{T},l}\right) + \vect{u}_{\perp}\trans\left(\theta_{\text{R},l}\right), \; -\left(1 + \cos\left(\Delta\theta_l\right)\right)d_{\text{R},\text{s},l}, \; 0, \; 0]\trans\in\mathbb{R}^5,\label{eq:J_NLOS dynamic direction tau theta}\\
				\vect{z}_{\theta, v, l} &=& [0, \; 0, \; -d_{\text{R},\text{s},l}, \; d_{\text{R},\text{s},l} \vect{u}\trans\left(\theta_{\text{R},l}\right)/\rho_l]\trans\in\mathbb{R}^5,\label{eq:J_NLOS dynamic direction theta v}\\
				\vect{z}_{\tau, v, l} &=& \vect{z}_{\tau, \theta, l} - \left(1 + \cos\left(\Delta\theta_l\right)\right)\vect{z}_{\theta, v, l}\label{eq:z_tau,v,l},
				\IEEEeqnarraynumspace
			\end{IEEEeqnarray}
			and
			\begin{IEEEeqnarray}{rCl}
				f_{\tau, \theta, l} &=& \frac{i_{\tau_l} i_{\theta_{\text{R},l}} i_{\theta_{\text{T},l}}}{\chi_l},\; f_{\tau, v, l} = \frac{i_{\tau_l} i_{v_l} i_{\theta_{\text{T},l}}}{\chi_l},\;f_{\theta, v, l} = \frac{i_{\theta_{\text{R},l}} i_{v_l}}{\chi_l}\left( i_{\tau_l}\left(1 + \cos\left(\Delta\theta_l\right)\right)^2 + i_{\theta_{\text{T},l}}\sin^2\left(\Delta\theta_l\right)\right),\nonumber\\
				\chi_l &=& \left(1 + \cos\left(\Delta\theta_l\right)\right)^2 i_{\tau_l}\left( i_{\theta_{\text{T},l}}  + i_{\theta_{\text{R},l}} + i_{v_l}  \right) + \sin^2\left(\Delta\theta_l\right)i_{\theta_{\text{T},l}} \left(i_{\theta_{\text{R},l}} + i_{v_l}\right),\nonumber\\
				i_{\tau_l} &=& \frac{\beta_l^2}{c^2}, \;\;\; i_{\theta_{\text{T},l}} = \frac{\omega_c^2 S_{\text{T}}(\tilde{\theta}_{\text{T},l})}{c^2 d_{\text{T},\text{s},l}^2}, \;\;\; i_{\theta_{\text{R},l}} = \frac{\omega_c^2 S_{\text{R}}(\tilde{\theta}_{\text{R},l})}{c^2 d_{\text{R},\text{s},l}^2}, \;\;\; i_{v_l} = \frac{\omega_c^2\rho_l^2 t_{\text{rms},l}^2}{c^2 d_{\text{R},\text{s},l}^2}.
				\IEEEeqnarraynumspace
				\label{eq:definitions of information quantities in dynamic localization}
			\end{IEEEeqnarray}
			The information loss due to the synchronization error is described by
			\begin{IEEEeqnarray}{rCl}
				\vect{J}_{\epsilon_{\text{clk}}}^{(\text{d})} &=&\vect{z}_{\epsilon_{\text{clk}}} \vect{z}_{\epsilon_{\text{clk}}}\trans/K_{\epsilon_{\text{clk}}}, \label{eq:J_coupl receiver no time-stamp dynamic}
				\IEEEeqnarraynumspace
			\end{IEEEeqnarray}
			where
			\begin{IEEEeqnarray}{rCl}
				\vect{z}_{\epsilon_{\text{clk}}} &=& \begin{cases}
					\frac{\beta_0^2}{c^2}\xi_0\vect{z}_{\tau_0} + \sum_{l=1}^L \left|h_l\right|^2 \sin\big(\Delta\theta_l\big) \xi_l\left( \left(f_{\tau, \theta, l} + f_{\tau, v, l}\right) \vect{z}_{\tau, \theta, l} + f_{\tau, v, l} \vect{z}_{\tau, v, l}\right), & \text{if LOS exists,}\\
					\sum_{l=0}^L \left|h_l\right|^2 \sin\big(\Delta\theta_l\big) \xi_l\left( f_{\tau, \theta, l} \vect{z}_{\tau, \theta, l} + f_{\tau, v, l} \vect{z}_{\tau, v, l}\right), & \text{otherwise,}
				\end{cases}, \nonumber\\
				K_{\epsilon_{\text{clk}}} &=& \begin{cases}
					\frac{1}{\delta_{\text{R},\text{T}}\sigma_{\text{clk}}^2} + \left|h_0\right|^2 \frac{\beta_0^2}{c^2}\xi_0 + \sum_{l'=1}^{L-1} \left|h_{l'}\right|^2 \xi_{l'} \sin^2\big(\Delta\theta_{l'}\big) \left(f_{\tau, \theta, l} + f_{\tau, v, l} \right), & \text{if LOS exists,}\\
					\frac{1}{\delta_{\text{R},\text{T}}\sigma_{\text{clk}}^2} + \sum_{l'=0}^{L-1} \left|h_{l'}\right|^2 \xi_{l'} \sin^2\big(\Delta\theta_{l'}\big) \left(f_{\tau, \theta, l} + f_{\tau, v, l} \right), & \text{otherwise.}\\
				\end{cases}.
				\label{eq:K_tau_s dynamic}\nonumber
				\IEEEeqnarraynumspace
			\end{IEEEeqnarray}
		\end{theorem}
		\begin{IEEEproof}
			Similar to the proof of Theorem~\ref{theor:receiver positioning without time stamps}.
		\end{IEEEproof}
		Some remarks on the theorem above should be made. 
		From \eqref{eq:v_LOS,v receiver dynamic}, we see that the Doppler shift of the LOS offers velocity information in the radial direction. In addition, it provides position information in the transverse direction, which corresponds to angle information,
		The intuition behind this is that, as an antenna array provides angle information by sampling the space at different locations \emph{simultaneously}, the movement of the receiver also provides angle information by allowing every single antenna element to sample the space at different locations \emph{over time}. The intensity of the position information is proportional to $\rho_{0}^2 t_{\text{rms},0}^2$. As explained in~\cite{HSZ+16}, $\rho_{0} t_{\text{rms},0}$ can be interpreted as the \emph{synthetic array aperture} for the LOS path, which is generated by the movement of the receiver.
		
		For the $l$-th single-bounce NLOS path the synthetic array aperture is $\rho_l t_{\text{rms},l}$. From \eqref{eq:J_NLOS receiver no time-stamp dynamic}-\eqref{eq:z_tau,v,l} we conclude that, from rank-$1$ Fisher information in $\mathbb{R}^3$ in the static scenario, the single-bounce NLOS paths now offer rank-$2$ information in $\mathbb{R}^5$. This results from the fact that $\vect{z}_{\tau, v, l}$ is a linear combination of $\vect{z}_{\tau, \theta, l}$ and $\vect{z}_{\theta, v, l}$. The EFIM from the $l$-th path is composed by the three rank-$1$ matrices determined by $\vect{z}_{\tau, \theta, l}, \vect{z}_{\tau, v, l}$ and $\vect{z}_{\theta, v, l}$. Only the first two matrices offer position information, in the same direction as in the static scenario. More specifically, the first matrix offers position and orientation information and the second matrix position and velocity information. The third matrix contains information only about the orientation and the velocity of the receiver. 
		
		In case the LOS path exists, at least one NLOS path is additionally required to have a full rank $\vect{J}_{\text{pov}}$, i.e. in order to be able to estimate the position, orientation and velocity. When only the LOS path is available, the transverse velocity cannot be estimated. Writing $\vect{v}_{\text{R}} = v_{\text{r}}\vect{u}\left(\theta_{\text{R},0}\right) + v_{\perp}\vect{u}\left(\theta_{\text{R},0}\right)$, where $v_r$ and $v_{\perp}$ are the radial and transverse velocity components, and plugging it into the dynamic signal model \eqref{eq:signal model dynamic}, \eqref{eq:dynamic channel definition}, we can see that only the radial velocity is observable. This could also be seen from \eqref{eq:v_LOS,v receiver dynamic}, where we observe that the LOS path offers velocity information only in the radial direction. Therefore, the LOS path allows us to estimate the position and the radial velocity $v_r$, but the NLOS paths enable us to obtain the missing orthogonal component $v_{\perp}$ of the velocity vector. This could be very important in the prediction step of position tracking algorithms. In the NLOS-only case three paths are required to obtain the position, orientation and velocity of the receiver, even in asynchronous operation. 
		
		Similar results, which are omitted for brevity, can be obtained in the case with a moving transmitter with unknown orientation and a static receiver \emph{(mobile transmitter localization)}.
		
	\section{Numerical Results}
		\label{sec:numerical results}
		In this section, we numerically evaluate the expressions derived above.  First, we study the effect of synchronization errors on the achievable positioning accuracy. Then, we investigate the impact of mobility on the PEB for different channel types (LOS-only, NLOS-only, LOS + NLOS) and different levels of synchronization error variance. Finally, we conduct a DL and UL positioning comparison with random beamformers.
		
		\subsection{Simulation parameters} 
		\label{sec:simulation parameters}
		For the channel model we follow~\cite{MWB+17}: the magnitude of the coefficient of the LOS path is $|h_0| = \lambda_c/(4\pi d_{\text{T},\text{R}})$, the magnitude of the NLOS paths, which are assumed to come from reflections, is $|h_l| = \Gamma_{\text{refl}}\lambda_c/(4\pi(d_{\text{T},\text{s}, l} + d_{\text{R},\text{s}, l}))$ with $\Gamma_{\text{refl}}=0.1$, and the phase of all the paths is uniformly distributed. 
		
		We consider a setup where the transmitter and the receiver are equipped with UCAs with $\lambda_c/2$-spaced elements, where $\lambda_c = c/f_{\text{c}}$ with $f_{\text{c}} = \SI{38}{\GHz}$. The orientation of the arrays is irrelevant as the SAAF of a UCA is independent of the angle. Unless otherwise specified, we set $N_{\text{T}} = N_{\text{R}} = 32$. and fix the position of the transmitter at $\vect{p}_{\text{T}} = [\SI{0}{\meter},\; \SI{0}{\meter}]\trans$, the position of the reflectors at $\vect{p}_{\text{s},1} = [\SI{10}{\meter},\; \SI{0}{\meter}]\trans$ and $\vect{p}_{\text{s},1} = [\SI{-8}{\meter},\; \SI{12}{\meter}]\trans$. The receiver lies in a $\SI{60}{\meter}\times \SI{60}{\meter}$ area with the transmitter at its center, i.e. $\SI{-30}{\meter} \leq p_{\text{R},x} \leq \SI{30}{\meter}$, $\SI{-30}{\meter} \leq p_{\text{R},y} \leq \SI{30}{\meter}$. 
		The sampling frequency is set to $F_{\text{s}} = 245.76$ MHz and the DFT size is $N = 1024$, resulting in a subcarrier spacing of $240$ kHz. One OFDM symbol is transmitted ($N_{\text{B}}=1$), where the set of occupied subcarriers is $\mathcal{P} = \left\{-297, -291, \ldots, 291, 297\right\}$ and $\gamma_p = 1/|\mathcal{P}|,\;\forall p\in\mathcal{P}$, with $|\mathcal{P}|=100$. The power per subcarrier is $\SI{0}{\dBm}$, thus $P_{\text{T}}=\SI{20}{\dBm}$. The noise spectral density is $N_0 = \SI{-170}{\dBm\per\hertz}$. 
		
		\subsection{The importance of synchronization}
		\label{sec:numerical results - effect of KTT}
		We now compare systems with or without synchronization errors for receiver localization. In Fig.~\ref{fig:bound over region, varying KTT} we plot the asymptotic PEB for a receiver with perfect synchronization (Fig.~\ref{fig:bound over region, varying KTT}(a)), with imperfect synchronization with $\sigma_{\text{clk}} = 0.5 T_{\text{s}} \approx \SI{2}{\nano\second}$ (Fig.~\ref{fig:bound over region, varying KTT}(b)), and without synchronization (Fig.~\ref{fig:bound over region, varying KTT}(c)). 
		\captionsetup[subfigure]{margin=0pt, singlelinecheck=true}
		\begin{figure}
			\centering
			\subfloat[Perfect synchornization.]{
				\includegraphics[scale=1]{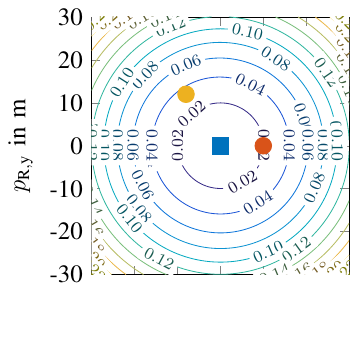}
			}%
			\subfloat[Imperfect synchronization: $\sigma_{\text{clk}} = 0.5 T_{\text{s}}\approx \SI{2}{\nano\second}$.]{
				\includegraphics[scale=1]{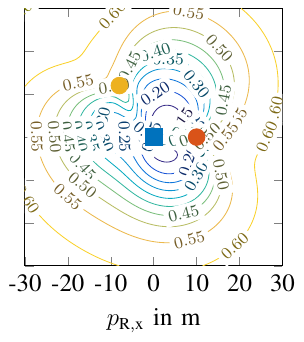}
			}
			\subfloat[No synchronization.]{
				\includegraphics[scale=1]{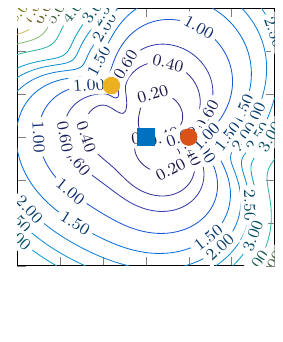}
			}
			\caption{PEB in m for receiver localization with $\vect{p}_{\text{T}} = [\SI{0}{\meter},\;\SI{0}{\meter}]\trans, \vect{p}_{\text{s},1} = [\SI{10}{\meter},\;\SI{0}{\meter}]\trans$ and $\vect{p}_{\text{s},2} = [\SI{-8}{\meter},\;\SI{12}{\meter}]\trans$ under different levels of synchronization error variance. The position of the transmitter is shown with a square and the positions of the reflectors with circles.
			}
			\label{fig:bound over region, varying KTT}
		\end{figure}
		Sub-meter positioning accuracy is attainable for all the considered receiver locations, when perfect synchronization or imperfect synchronization with $\sigma_{\text{clk}} = 0.5 T_{\text{s}}$ is available. However, in asynchronous operation (Fig.~\ref{fig:bound over region, varying KTT}(c)), there is significant accuracy degradation because the positioning information from the LOS is no longer sufficient and information from the much weaker NLOS paths is required to compute the position. We can see in Fig. \ref{fig:bound over region, varying KTT}(b) that clock synchronization errors with standard deviation as small as $0.5 T_{\text{s}}\approx\SI{2}{\nano\second}$ can severely degrade the achievable positioning accuracy, compared to the case with perfect synchronization.  
		
		We observe in Fig.~\ref{fig:bound over region, varying KTT} that with perfect synchronization the PEB depends only on the distance from the transmitter, as the contour lines are concentric circles with the transmitter at their center. This is attributed to the fact that the distance and angle information from the LOS path are dominant. With imperfect or no synchronization, the distance information from the LOS path is significantly reduced or vanished. The information offered by the NLOS paths, especially the information in the radial direction, allows the receiver to resolve the synchronization error and hence compute its distance from the transmitter more accurately. Since the direction and the intensity of the information from each NLOS path depends on the relative position of the receiver with respect to the reflector (see \eqref{eq:J_NLOS intensity of information} and \eqref{eq:J_NLOS direction of information}), the contour lines in Figs.~\ref{fig:bound over region, varying KTT}(b) and \ref{fig:bound over region, varying KTT}(c) are influenced not only by the position of the transmitter but also from the position of the reflectors. A characteristic example is when the receiver lies in the area behind a reflector. In this case $\vect{u}_{\perp}\left(\theta_{\text{T},l}\right) \approx -\vect{u}_{\perp}\left(\theta_{\text{R},l}\right)$ and, thus, from \eqref{eq:J_NLOS direction of information} we can conclude that the information offered by the corresponding path is significantly reduced. This explains the curvature of the contour lines in the regions behind the reflectors.
		
\begin{figure}
			\centering
			\includegraphics[scale=1]{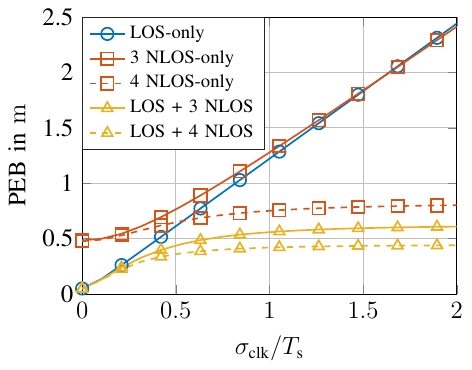}
			\caption{PEB vs $\sigma_{\text{clk}}$ for LOS and NLOS static receiver localization with $\vect{p}_{\text{T}} = [\SI{0}{\meter},\;\SI{0}{\meter}],\;\vect{p}_{\text{R}} = [\SI{18}{\meter},\;\SI{3}{\meter}]\trans, \;\vect{p}_{\text{s},1} = [\SI{10}{\meter},\;\SI{0}{\meter}]\trans$, $\vect{p}_{\text{s},2} = [\SI{-8}{\meter},\;\SI{12}{\meter}]\trans,\;\vect{p}_{\text{s},3} = [\SI{12}{\meter},\;\SI{8}{\meter}]\trans,\;\vect{p}_{\text{s},4} = [\SI{7}{\meter},\;\SI{-6}{\meter}]\trans$.}
			\label{fig:PEB vs sigma_clk}
		\end{figure}

		In order to further study the effect of synchronization on positioning accuracy, in Fig.~\ref{fig:PEB vs sigma_clk} we plot the PEB as a function of $\sigma_{\text{clk}}/T_{\text{s}}$ with the receiver's position fixed at $\vect{p}_{\text{R}} = [\SI{18}{\meter},\;\SI{3}{\meter}]\trans$.
		To also evaluate the performance of NLOS-only positioning, which requires at least 4 paths in the asynchronous case, we add extra reflectors at $\vect{p}_{\text{s},3} = [\SI{12}{\meter},\;\SI{8}{\meter}]\trans$ and $\vect{p}_{\text{s},4} = [\SI{7}{\meter},\;\SI{-6}{\meter}]\trans$. 
		The following 5 cases are considered: (i) \emph{LOS-only}: only the LOS path is available; (ii) \emph{3 NLOS-only}: only 3 NLOS paths (corresponding to $\vect{p}_{\text{s},1}, \vect{p}_{\text{s},2}$ and $\vect{p}_{\text{s},3}$) are available; (iii) \emph{LOS + 3 NLOS}: the LOS and the 3 NLOS paths are available; (iv) \emph{4 NLOS-only}: only 4 NLOS paths are available; (v) \emph{LOS + 4 NLOS}: the LOS and the 4 NLOS paths are available.
		In the LOS-only case and in the NLOS-only case with 3 NLOS paths, the accuracy is significantly degraded with increasing $\sigma_{\text{clk}}$, as the performance of these cases is lower-bounded by the standard deviation of the ranging error caused by the synchronization error, which is equal to $\sigma_{\text{clk}}c$ (see also the discussion after Theorem~\ref{theor:receiver positioning without time stamps}). On the other hand, with 4 NLOS paths, the PEB of NLOS-only positioning saturates with increasing $\sigma_{\text{clk}}$, as the synchronization error can be resolved using the paths. Similarly, when both LOS and NLOS paths are present the PEB saturates, with the saturation value improving with increasing number of paths. 
		
		\subsection{The effect of Doppler shifts}
		We consider the setup described in Section~\ref{sec:simulation parameters}, but the receiver is now moving with velocity $\vect{v}_{\text{R}} = \vect{u}(\pi/2)\SI{60}{\kilo\meter\per\hour}$. Instead of 1 OFDM symbol, we now transmit $N_{\text{B}} = 32$ OFDM symbols, such that the synthetic array aperture is comparable to the physical aperture of the antenna arrays. The duration of the CP is $72 T_{\text{s}}\approx 0.07NT_{\text{s}}$. We keep the total transmit power the same as in the static scenario, i.e. $N_{\text{B}} P_{\text{T}} = \SI{20}{\dBm}$, so as to observe the effect of the receiver's movement for the same receive SNR. 
		\captionsetup[subfigure]{margin=0pt, singlelinecheck=true}
		\begin{figure}
			\centering
			\includegraphics[scale=1]{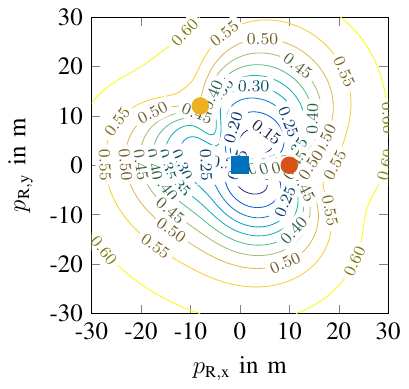}
			\caption{PEB in m for receiver localization with $\vect{p}_{\text{T}} = [\SI{0}{\meter},\;\SI{0}{\meter}]\trans, \vect{p}_{\text{s},1} = [\SI{10}{\meter},\;\SI{0}{\meter}]\trans$ and $\vect{p}_{\text{s},2} = [\SI{-8}{\meter},\;\SI{12}{\meter}]\trans$, receiver velocity $\vect{v}_{\text{R}} = \vect{u}(\pi/2)\SI{60}{\kilo\meter\per\hour}$, $\sigma_{\text{clk}} = 0.5 T_{\text{s}}\approx \SI{2}{\nano\second}$. The position of the transmitter is shown with a square and the positions of the reflectors with circles.}
			\label{fig:bound over region imperfect KTT dynamic}
		\end{figure}
		We can see in Fig.~\ref{fig:bound over region imperfect KTT dynamic} that, compared to the static case, no considerable improvement is achieved except for the areas behind the reflectors. As explained in the discussion of Fig.~\ref{fig:bound over region, varying KTT}, in these regions, 
		the movement-independent part of the information from the corresponding NLOS path (see~\eqref{eq:J_NLOS receiver no time-stamp dynamic},\eqref{eq:J_NLOS dynamic direction tau theta}) is reduced. Therefore, the additional information offered by the Doppler shift (see~\eqref{eq:J_NLOS receiver no time-stamp dynamic}, \eqref{eq:J_NLOS dynamic direction theta v}, \eqref{eq:z_tau,v,l}) becomes important.
		
\begin{figure}
			\centering
			\includegraphics[scale=1]{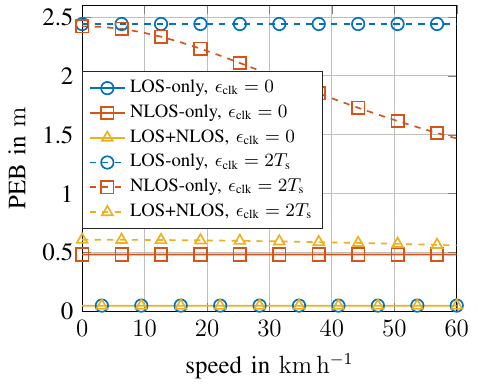}
			\caption{PEB vs speed for LOS and NLOS receiver localization for perfect ($\sigma_{\text{clk}} = 0$) and imperfect synchronization ($\sigma_{\text{clk}} = 2 T_{\text{s}}$) with $\vect{p}_{\text{T}} = [\SI{0}{\meter},\;\SI{0}{\meter}],\;\vect{p}_{\text{R}} = [\SI{18}{\meter},\;\SI{3}{\meter}]\trans, \;\vect{p}_{\text{s},1} = [\SI{10}{\meter},\;\SI{0}{\meter}]\trans$ and $\vect{p}_{\text{s},2} = [\SI{-8}{\meter},\;\SI{12}{\meter}]\trans,\;\vect{p}_{\text{s},3} = [\SI{12}{\meter},\;\SI{8}{\meter}]\trans$, velocity direction $\vect{v}_{\text{R}}/\left\|\vect{v}_{\text{R}}\right\|_2 = \vect{u}(\pi/2)$.}
			\label{fig:PEB vs velocity}
		\end{figure}

		To gain further insight on the effect of velocity, in Fig.~\ref{fig:PEB vs velocity} we plot the PEB as a function of the speed, i.e. $\left\|\vect{v}_{\text{R}}\right\|_2$, for a system with perfect ($\sigma_{\text{clk}}=0$) and imperfect ($\sigma_{\text{clk}}=2 T_{\text{s}}$) synchronization.
		Again, we fix the receiver's position at $\vect{p}_{\text{R}} = [\SI{18}{\meter},\;\SI{3}{\meter}]\trans$ and add an extra reflector at $\vect{p}_{\text{s},3} = [\SI{12}{\meter},\;\SI{8}{\meter}]\trans$. We consider three different cases: (i) \emph{LOS-only}: only the LOS path is available; (ii) \emph{NLOS-only}: only the 3 NLOS paths are available; (iii) \emph{LOS+NLOS}: the LOS and the 3 NLOS paths are available. A few interesting observations can be made from Fig.~\ref{fig:PEB vs velocity}. In the perfectly synchronized system ($\sigma_{\text{clk}} = 0$) the position information offered by the LOS path is dominant and the information from the NLOS path does not improve the PEB (LOS-only and LOS+NLOS curves coincide). In addition, the movement of the receiver does not improve the positioning accuracy in any of the three considered cases.
		In the imperfectly synchronized system ($\sigma_{\text{clk}} = 2T_{\text{s}}$), the accuracy of the LOS-only case is lower bounded by the standard deviation $\sigma_{\text{clk}}c \approx \SI{2.44}{\meter}$ of the ranging error induced by the synchronization error. From the LOS+NLOS curves, we first observe that the information from the NLOS paths is very important as it allows the receiver to resolve the synchronization error, as discussed before.
		Similarly to Fig.~\ref{fig:bound over region imperfect KTT dynamic}, we can see that the movement of the receiver has a small impact on the PEB, when the LOS path is present. On the contrary, in the NLOS-only case, mobility has a strong impact, with the PEB decreasing significantly with increasing speed. This can be understood by comparing \eqref{eq:J_NLOS receiver no time-stamp}-\eqref{eq:J_NLOS direction of information} with \eqref{eq:J_NLOS receiver no time-stamp dynamic}-\eqref{eq:z_tau,v,l}: the additional sources of position and orientation information (2nd and 3rd matrices in \eqref{eq:J_NLOS receiver no time-stamp dynamic}), whose intensities are proportional to $\rho_l^2$, allow the receiver to better resolve the synchronization error, while in the static scenario its accuracy is bounded by $\sigma_{\text{clk}}c$.
		
		\subsection{Comparison of Uplink and Downlink Positioning}
\begin{figure}
			\centering
			\includegraphics[scale=1]{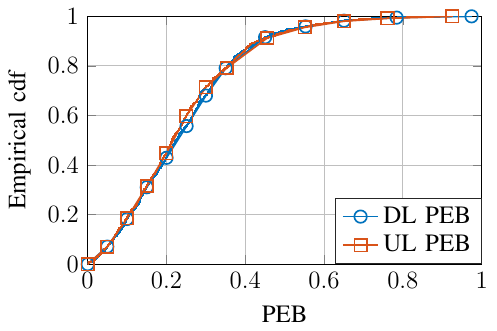}
			\caption{DL and UL positioning comparison with random beamforming vectors for transmission and reception.}
			\label{fig:DL UL with beamforming}
		\end{figure}

		In Section~\ref{sec:downlink and uplink positioning}, we studied the relation of the PEB in the DL and the UL when the same set of orthonormal beamforming vectors are used by each device for transmission and reception. We found that, under the same receive SNR, DL and UL positioning are equivalent. Here, we conduct a numerical comparison of DL and UL positioning when the devices employ random beamforming vectors for transmission and reception. We consider the same setup as before, with perfect synchronization, but we now consider LOS-only propagation and set $M_{\text{BS}} = 16$, which is $M_{\text{T}}$ in the DL and $M_{\text{R}}$ in the UL. Similarly, we set $M_{\text{UE}} = 4$, which is $M_{\text{R}}$ in the DL and $M_{\text{T}}$ in the UL. At each considered UE position, the entries of all beamforming matrices are unimodular with random phase uniformly distributed in $[-\pi,\pi)$.
		The available transmit power is uniformly allocated to the beamforming vectors and the subcarriers, i.e. $\sum_{b = 0}^{N_{\text{B}} - 1} \vect{s}_b[p]\vect{s}_b\herm[p] = N_{\text{B}}P_{\text{T}}/(M_{\text{T}}|\mathcal{P}|) \vect{I}_{M_{\text{T}}}$. Motivated by our findings about the relation between DL and UL positioning in Section~\ref{sec:downlink and uplink positioning}, we set $\sigma_{\eta,\text{UE}}^2 = 2\sigma_{\eta,\text{BS}}^2$ and $P_{\text{UE}} = P_{\text{BS}}\frac{\sigma_{\eta,\text{BS}}^2}{\sigma_{\eta,\text{UE}}^2}\frac{M_{\text{UE}}}{M_{\text{BS}}}$, which leads to the same receive SNR in the DL and the UL, with $N_B P_{\text{BS}} = \SI{20}{\dBm}$ and the noise spectral density of the BS $N_{0,\text{BS}}=\SI{-170}{\dBm\per\hertz}$. In Fig.~\ref{fig:DL UL with beamforming}, using the expressions for the exact CRLB \eqref{eq:J_phi definition}-\eqref{eq:entries of channel parameter FIM}, \eqref{eq:J_po_definition}, \eqref{eq:PEB definition}, we plot the cumulative distribution function (cdf) for the PEB in the DL and UL for $\alpha_{\text{UE}} = \ang{180}$. 
		We can observe in Fig.~\ref{fig:DL UL with beamforming} that, under the same receive SNR, the DL and UL PEB are approximately equal, similarly to the case of orthonormal beamforming. In general, the link with the highest receive SNR will provide the best performance.

	\section{Conclusion}
		\label{sec:conclusion}
		We derived asymptotic expressions for the Fisher information on position, orientation and velocity for transmitter and receiver localization in a MIMO-OFDM setup, considering synchronization errors and mobility of the transmitter or receiver. We studied the impact of synchronization errors and found that even small synchronization errors can significantly degrade the achievable positioning accuracy. Through analytical and numerical results we showed that NLOS paths can help resolve the clock offset and drastically improve the PEB. In addition, we investigated the effect of mobility and proved that each single-bounce NLOS path offers rank-2 information on position, orientation and velocity. Our numerical results showed that in the presence of the LOS path mobility has no significant impact on the achievable positioning accuracy. However, for NLOS-only positioning, it can provide valuable information that can drastically improve the positioning accuracy of an imperfectly synchronized system. Finally, we compared DL and UL positioning with orthonormal or random beamforming and found that they are equivalent when the DL and UL receive SNR are equal.

	\appendices

		\section{Derivatives of \texorpdfstring{${\vect{m}}_b[p]$}{mb[p]} w.r.t. the parameters in the channel parameter vector}
			\label{sec:appendix - derivative of m_b[p] wrt to phi}
			We provide the required derivatives of ${\vect{m}}_b[p]$ for the dynamic scenario; the required derivatives for the static scenario can be obtained from them by setting $\vect{v}_{\text{R}} = \vect{0}$ (or equivalently $v_l=0,\;\forall l$), bearing in mind that $\lim_{v_l\rightarrow 0} Q(\tilde{\Phi}_{p,p,l}) = 1$ and $\lim_{v_l\rightarrow 0} Q(\tilde{\Phi}_{p,q,l}) = 0, \;p\neq q$, where $\tilde{\Phi}_{p,q,l} = \left( \omega_p - \omega_q - \left(\omega_c + \omega_q\right)v_{l}/c \right)T_{\text{s}}/2$. Setting $\tilde{\Psi}_{q,l,b} = \omega_q \left(\tau_l + \epsilon_{\text{clk}}\right) - \frac{\omega_c + \omega_q}{c}v_l b M T_{\text{s}}$, we compute the required derivatives as 
			\begin{IEEEeqnarray}{rCl}
				\partial {\vect{m}}_b[p]/\partial \epsilon_{\text{clk}} &=& -\jj \sum\nolimits_{q\in\mathcal{P}} \omega_q {g}[q] \sum\nolimits_{l=0}^{L-1} h_l \e^{-\jj \tilde{\Psi}_{q,l,b}} Q(\tilde{\Phi}_{p,q,l}) \vect{F}_{\text{R}}\herm\vect{a}_{\text{R},q}(\tilde{\theta}_{\text{R},l})\vect{a}_{\text{T},q}\trans (\tilde{\theta}_{\text{T},l}){\vect{x}}_b\left[q\right],\label{eq:derivative epsilon_clk}\\
				\partial {\vect{m}}_b[p]/\partial \tau_l &=& -\jj \sum\nolimits_{q\in\mathcal{P}} \omega_q {g}[q]  h_l \e^{-\jj \tilde{\Psi}_{q,l,b}} Q(\tilde{\Phi}_{p,q,l})\vect{F}_{\text{R}}\herm\vect{a}_{\text{R},q}(\tilde{\theta}_{\text{R},l})\vect{a}_{\text{T},q}\trans (\tilde{\theta}_{\text{T},l}){\vect{x}}_b\left[q\right],\\
				\partial {\vect{m}}_b[p]/\partial \tilde{\theta}_{\text{T},l} &=& \sum\nolimits_{q\in\mathcal{P}} {g}[q]  h_l \e^{-\jj \tilde{\Psi}_{q,l,b}} Q(\tilde{\Phi}_{p,q,l}) \vect{F}_{\text{R}}\herm\vect{a}_{\text{R},q}(\tilde{\theta}_{\text{R},l})\vect{a}_{\text{T},q}\trans (\tilde{\theta}_{\text{T},l})\vect{D}_{\text{T},q}\trans (\tilde{\theta}_{\text{T},l}){\vect{x}}_b\left[q\right],\\
				\partial {\vect{m}}_b[p]/\partial \tilde{\theta}_{\text{R},l} &=& \sum\nolimits_{q\in\mathcal{P}} {g}[q]  h_l \e^{-\jj \tilde{\Psi}_{q,l,b}} Q(\tilde{\Phi}_{p,q,l})\vect{F}_{\text{R}}\herm\vect{D}_{\text{R},q} (\tilde{\theta}_{\text{R},l})\vect{a}_{\text{R},q}(\tilde{\theta}_{\text{R},l})\vect{a}_{\text{T},q}\trans (\tilde{\theta}_{\text{T},l}){\vect{x}}_b\left[q\right],\label{eq:derivative theta_R}\\
				\partial {\vect{m}}_b[p]/\partial h_{l,\Re} &=& \sum\nolimits_{q\in\mathcal{P}} {g}[q] \e^{-\jj \tilde{\Psi}_{q,l,b}} Q(\tilde{\Phi}_{p,q,l})\vect{F}_{\text{R}}\herm\vect{a}_{\text{R},q}(\tilde{\theta}_{\text{R},l})\vect{a}_{\text{T},q}\trans (\tilde{\theta}_{\text{T},l}){\vect{x}}_b\left[q\right],\label{eq:derivative h_R}\\
				\partial {\vect{m}}_b[p]/\partial h_{l,\Im} &=& \jj \partial {\vect{m}}_b[p]/\partial h_{l,\Re}\label{eq:derivative h_I},\\
				\partial {\vect{m}}_b[p]/\partial v_l &=& \sum\nolimits_{q\in\mathcal{P}} {g}[q]  h_l \e^{-\jj \tilde{\Psi}_{q,l,b}} Q(\tilde{\Phi}_{p,q,l}) V_{b,p,q,l}\vect{F}_{\text{R}}\herm \vect{a}_{\text{R},q}(\tilde{\theta}_{\text{R},l})\vect{a}_{\text{T},q}\trans (\tilde{\theta}_{\text{T},l}){\vect{x}}_b\left[q\right],
				\IEEEeqnarraynumspace
			\end{IEEEeqnarray}
			where $\vect{D}_{\text{T},q} (\tilde{\theta}_{\text{T},l})$ is a diagonal matrix with $[\vect{D}_{\text{T},q} (\tilde{\theta}_{\text{T},l})]_{j,j} = -\jj \frac{\omega_c + \omega_q}{c} d_{\text{T},j} \vect{u}_{\perp}\trans(\tilde{\theta}_{\text{T},l}) \vect{u}(\psi_{\text{T},j})$ and $\vect{D}_{\text{R},p} (\tilde{\theta}_{\text{R},l})$ is defined accordingly. Also
			\begin{IEEEeqnarray}{rCl}
				V_{b,p,q,l} &=& \big(\cot(\tilde{\Phi}_{p,q,l}) - N\cot(N\tilde{\Phi}_{p,q,l})\big)T_{\text{s}}/2  + \jj (bM + (N-1)/2)T_{\text{s}}.
				\IEEEeqnarraynumspace
			\end{IEEEeqnarray}

	\section{Entries of Fisher Information Transformation Matrices}
		\label{sec:appendix - entries of Fisher information transformation matrices}
		The entries of $\vect{T}_{\text{T}}$ ($\vect{T}_{\text{R}}$) corresponding to identical parameters in $\vect{\phi}$ and $\tilde{\vect{\phi}}_{\text{T}}$ ($\tilde{\vect{\phi}}_{\text{R}}$) are equal to $1$, e.g. $\partial \epsilon_{\text{clk}}/\partial\epsilon_{\text{clk}}=1$. 
		Setting $d_{\text{T},\text{R}} = \left\|\vect{p}_{\text{R}} - \vect{p}_{\text{T}}\right\|_2$, $d_{\text{T},\text{s},l} = \left\|\vect{p}_{\text{s},l} - \vect{p}_{\text{T}}\right\|_2$ and
		$d_{\text{R},\text{s},l} = \left\|\vect{p}_{\text{R}} - \vect{p}_{\text{s},l}\right\|_2$, the rest non-zero entries are computed as follows:\newline
		\begin{minipage}{0.5\linewidth}
			\noindent
			\begin{IEEEeqnarray}{rCl}
				\partial \tau_{l}/\partial \vect{p}_{\text{R}} &=& - \vect{u}\left(\theta_{\text{R},l}\right)/c,\label{eq:T tau p_R}\\
				\partial \tilde{\theta}_{\text{T},l}/\partial \vect{p}_{\text{R}} &=&  \begin{cases}
					\frac{\vect{u}_{\perp}\left(\theta_{\text{R},l}\right)}{d_{\text{T},\text{R}}}, \; l\text{-th path is LOS},\\
					\vect{0}, \; \text{otherwise},\\
				\end{cases}\\
				\partial \tilde{\theta}_{\text{R},l}/\partial \vect{p}_{\text{R}} &=&  \begin{cases}
					\frac{\vect{u}_{\perp}\left(\theta_{\text{R},l}\right)}{d_{\text{T},\text{R}}}, \; l\text{-th path is LOS},\\
					\frac{\vect{u}_{\perp}\left(\theta_{\text{R},l}\right)}{d_{\text{R},\text{s},l}}, \; \text{otherwise},\\
				\end{cases}\\
				\partial \tilde{\theta}_{\text{R},l}/\partial \alpha_{\text{R}} &=& -1,\label{eq:T theta_R alpha_R}
				\IEEEeqnarraynumspace
			\end{IEEEeqnarray}
		\end{minipage}%
		\begin{minipage}{0.5\linewidth}
			\vspace*{-0.4cm}
			\begin{IEEEeqnarray}{rCl}
				\partial \tau_{l}/\partial \vect{p}_{\text{T}} &=& - \vect{u}\left(\theta_{\text{T},l}\right)/c,\\
				\partial \tilde{\theta}_{\text{T},l}/\partial \vect{p}_{\text{T}} &=&  \begin{cases}
					\frac{\vect{u}_{\perp}\left(\theta_{\text{T},l}\right)}{d_{\text{T},\text{R}}}, \; l\text{-th path is LOS},\\
					\frac{\vect{u}_{\perp}\left(\theta_{\text{T},l}\right)}{d_{\text{T},\text{s},l}}, \; \text{otherwise},\\
				\end{cases}\\
				\partial \tilde{\theta}_{\text{R},l}/\partial \vect{p}_{\text{T}} &=&  \begin{cases}
					\frac{\vect{u}_{\perp}\left(\theta_{\text{T},l}\right)}{d_{\text{T},\text{R}}}, \; l\text{-th path is LOS},\\
					\vect{0}, \; \text{otherwise},\\
				\end{cases}\\
				\partial \tilde{\theta}_{\text{T},l}/\partial \alpha_{\text{T}} &=& -1,
				\IEEEeqnarraynumspace
			\end{IEEEeqnarray}
		\end{minipage}
		\begin{IEEEeqnarray}{rCl}
			\partial \tau_{l}/{\partial \vect{p}_{\text{s},l}} &=& (\vect{u}\left(\theta_{\text{T},l}\right) + \vect{u}\left(\theta_{\text{R},l}\right))/c,\\
			\partial \tilde{\theta}_{\text{T},l}/{\partial \vect{p}_{\text{s},l}} &=& -\vect{u}_{\perp}\left(\theta_{\text{T},l}\right)/d_{\text{T},\text{s},l}, \\
			\partial \tilde{\theta}_{\text{R},l}\partial \vect{p}_{\text{s},l} &=& -\vect{u}_{\perp}\left(\theta_{\text{R},l}\right)/d_{\text{R},\text{s},l}.
			\IEEEeqnarraynumspace
		\end{IEEEeqnarray}
		For a mobile receiver
		\begin{IEEEeqnarray}{rCl}
			\partial v_l/\partial \vect{p}_{\text{R}} &=& -\rho_l \vect{u}_{\perp}\left(\theta_{\text{R},l}\right) \begin{cases}
				1/d_{\text{T},\text{R}}, & l\text{-th path is LOS},\\
				1/d_{\text{R},\text{s},l}, & \text{otherwise},\\
			\end{cases}\\
			\partial v_l/\partial \vect{v}_{\text{R}} &=& \vect{u}\left(\theta_{\text{R},l}\right),\\
			\partial v_l/\partial \vect{p}_{\text{s},l} &=& \rho_l \vect{u}_{\perp}\left(\theta_{\text{R},l}\right)/d_{\text{R},\text{s},l},
			\IEEEeqnarraynumspace
		\end{IEEEeqnarray}
		where $\rho_l = \vect{v}_{\text{R}}\trans\vect{u}_{\perp}\left(\theta_{\text{R},l}\right)$.	Similar expressions can be obtained for a mobile transmitter.
		
	\section{Proof of Theorem~\ref{theor:receiver positioning without time stamps}}
		\label{sec:proof of theorem for receiver positioning without time-stamps}
		We first have to compute the asymptotic expressions for the entries of the channel parameters FIM $\vect{J}_{\vect{\phi}}$. 
		In order to facilitate readability and understanding, instead of referring to the entries of $\vect{J}_{\vect{\phi}}^{(\text{o})}$ by their position in the matrix, e.g. $\big[\vect{J}_{\vect{\phi}}^{(\text{o})}\big]_{1,3}$, we refer to them by using as indices the variables they correspond to, e.g. $J_{\epsilon_{\text{clk}}\tilde{\theta}_{\text{T},0}}$. Using \eqref{eq:fav_prop asymptotic orthogonality}, \eqref{eq:expectation of entries of channel parameter FIM} and \eqref{eq:derivative epsilon_clk}-\eqref{eq:derivative h_I} and the assumptions 1-3 of Section~\ref{sec:geometric interpretation of Fisher information}, we find that
		
		\noindent\begin{minipage}[t]{0.47\linewidth}
			\vspace*{-0.5cm}
			\begin{IEEEeqnarray}{rCl}
				J_{\epsilon_{\text{clk}}\epsilon_{\text{clk}}} &\rightarrow& \delta_{\text{R},\text{T}} \sum_{p\in\mathcal{P}}\gamma_p \omega_p^2 \sum_{l=0}^{L-1}\left|h_l\right|^2,\label{eq:J_tau_s}\\
				J_{\tau_l\tau_l} &=&J_{\epsilon_{\text{clk}}\tau_l} \rightarrow \delta_{\text{R},\text{T}}  \sum_{p\in\mathcal{P}}\gamma_p \omega_p^2\left|h_l\right|^2, \label{eq:J_Deltatau_l}\\
				J_{\tilde{\theta}_{\text{T},l}\tilde{\theta}_{\text{T},l}} &=& \delta_{\text{R},\text{T}} \left|h_l\right|^2 \left(\bar{\omega}_c/c\right)^2 S_{\text{T}}(\tilde{\theta}_{\text{T},l}), \label{eq:J_theta_Tl}\\
				J_{\tilde{\theta}_{\text{R},l}\tilde{\theta}_{\text{R},l}} &=& \delta_{\text{R},\text{T}} \left|h_l\right|^2 \left(\bar{\omega}_c/c\right)^2 S_{\text{R}}(\tilde{\theta}_{\text{R},l}), \label{eq:J_theta_Rl}
				\IEEEeqnarraynumspace
			\end{IEEEeqnarray}
		\end{minipage}
		\begin{minipage}[t]{0.5\linewidth}
			\vspace*{-0.5cm}
			\begin{IEEEeqnarray}{rCl}
				J_{h_{l,\Re}h_{l,\Re}} &=& J_{h_{l,\Im}h_{l,\Im}} = \delta_{\text{R},\text{T}},\\
				J_{\epsilon_{\text{clk}} h_{l,\Re}} &=& J_{\tau_l h_{l,\Re}} \rightarrow \delta_{\text{R},\text{T}} h_{l,\Im}\sum_{p\in\mathcal{P}}\gamma_p\omega_p,\\
				J_{\epsilon_{\text{clk}} h_{l,\Im}} &=& J_{\tau_l h_{l,\Im}} \rightarrow -\delta_{\text{R},\text{T}}  h_{l,\Re}\sum_{p\in\mathcal{P}}\gamma_p\omega_p,
				\IEEEeqnarraynumspace
			\end{IEEEeqnarray}
		\end{minipage}
		\vspace*{0.4cm}
		
		\noindent where with some abuse of notation the equality sign is used to denote that the asymptotic values are equal. Using the notion of the EFIM, we can show that as far as position and orientation information is concerned, we can exclude the channel gains from the parameter vectors, as long as we include the uncertainty they introduce to the model by replacing the time-related FIM entries $J_{\epsilon_{\text{clk}}\epsilon_{\text{clk}}},  \left\{J_{\tau_l\tau_l}, J_{\epsilon_{\text{clk}}\tau_l}  \right\}_{l=0}^{L-1}$ with $J_{\epsilon_{\text{clk}}\epsilon_{\text{clk}}}',  \left\{J_{\tau_l\tau_l}', J_{\epsilon_{\text{clk}}\tau_l}'  \right\}_{l=0}^{L-1}$, where $J_{\epsilon_{\text{clk}}\epsilon_{\text{clk}}}' \rightarrow \delta_{\text{R},\text{T}} \beta^2 \sum_{l=0}^{L-1}\left|h_l\right|^2$ and
		$J_{\tau_l\tau_l}' =J_{\epsilon_{\text{clk}}\tau_l}' \rightarrow \delta_{\text{R},\text{T}} \beta^2\left|h_l\right|^2$.
		Using the favorable propagation conditions \eqref{eq:fav_prop asymptotic orthogonality}-\eqref{eq:fav_prop second derivative} and following the Bachmann-Landau asymptotic notations\footnote{$f(x) = o(g(x))$ means that $\lim_{x\rightarrow\infty}\left|\frac{f(x)}{g(x)}\right| = 0$ and $f(x) = \Theta(g(x))$ means that $0<\liminf_{x\rightarrow\infty}\left|\frac{f(x)}{g(x)}\right| \leq \limsup_{x\rightarrow\infty}\left|\frac{f(x)}{g(x)}\right| < \infty$}~\cite{Cor09}, we can show that all the diagonal entries of the FIM and $J_{\epsilon_{\text{clk}}\tau_l}$ are $\Theta(x^3)$, where $x\in N_{\text{T}}, N_{\text{R}}, F_{\text{s}}$, e.g. $J_{\tau_l\tau_l}=\Theta(N_{\text{R}}F_{\text{s}}^2)$, and the rest of the entries are $o(x^3)$, which means they can be ignored in the computation of the asymptotic position and orientation EFIM. For brevity, we only present the derivation for the case when the LOS path is available. In order to make the expressions more compact, when the same variable is used in both indices we write it only once; e.g. we write $J_{\epsilon_{\text{clk}}}'$ instead of $J_{\epsilon_{\text{clk}}\epsilon_{\text{clk}}}'$. 
		Using the expressions from Appendix~\ref{sec:appendix - entries of Fisher information transformation matrices} for the entries of the Fisher information transformation matrices, we compute
		\begin{IEEEeqnarray}{rCl}
			\vect{T}_{\text{po}} \vect{J}_{\vect{\phi}} \vect{T}_{\text{po}}\trans &=& 
			\frac{J_{\tau_0}'}{c^2} \vect{z}_{\tau_0}\vect{z}_{\tau_0}\trans + \frac{J_{\tilde{\theta}_{\text{T},0}}}{d_{\text{T},\text{R}}^2} \vect{z}_{\theta_{\text{T},0}}\vect{z}_{\theta_{\text{T},0}}\trans + \frac{J_{\tilde{\theta}_{\text{R},0}}}{d_{\text{T},\text{R}}^2} \vect{z}_{\theta_{\text{R},0}}\vect{z}_{\theta_{\text{R},0}}\trans + \sum_{l=1}^{L-1}\hspace*{-0.05cm}\left(\frac{J_{\tau_l}'}{c^2} \vect{z}_{\tau_l}\vect{z}_{\tau_l}\trans + \frac{J_{\tilde{\theta}_{\text{R},l}}}{d_{\text{R},\text{s},l}^2} \vect{z}_{\theta_{\text{R},l}}\vect{z}_{\theta_{\text{R},l}}\trans\right)\hspace*{-0.1cm},\label{eq:T_p J_phi T_p}\hspace*{-0.1cm}
			\IEEEeqnarraynumspace
		\end{IEEEeqnarray}
		where $\vect{z}_{\tau_l} \hspace*{-0.02cm}= \hspace*{-0.02cm}[
		- \vect{u}\trans\hspace*{-0.02cm}\left(\theta_{\text{R},l}\right), \; 0]\trans$ and $\vect{z}_{\theta_{\text{R},l}} \hspace*{-0.02cm}= \hspace*{-0.02cm}[\vect{u}_{\perp}\trans\hspace*{-0.02cm}\left(\theta_{\text{R},l}\right), \; -d_{\text{R},\text{s},l}]\trans$. We also write 
		\begin{IEEEeqnarray}{rCl}
			\vect{T}_{\text{np}} \vect{J}_{\vect{\phi}}\vect{T}_{\text{np}}\trans &=& \begin{bmatrix}
				J_{\epsilon_{\text{clk}}}' + \frac{1}{\sigma_{\text{clk}}^2} & \vect{b}\trans\\
				\vect{b} & \vect{C}
			\end{bmatrix},\label{eq:T_np J_phi T_np}
		\end{IEEEeqnarray}
		where $\vect{b} = \left[
		J_{\tau_1}'\vect{\nu}_1\trans/c, \ldots,  J_{\Delta\tau_{L-1}}'\vect{\nu}_{L-1}\trans/c\right]\trans$, 
		$\vect{\nu}_l = \vect{u}\left(\theta_{\text{T},l}\right) + \vect{u}\left(\theta_{\text{R},l}\right)$
		and $\vect{C}$ is a block diagonal matrix with the following $2\times 2$ matrices on its diagonal:
		\begin{IEEEeqnarray}{rCl}
			\vect{J}_{s,l} &=& \frac{J_{\tau_l}'}{c^2}\vect{\nu}_l\vect{\nu}_l\trans + \frac{J_{\tilde{\theta}_{\text{T},l}}}{d_{\text{T},\text{s},l}^2}\vect{u}_{\perp}\left(\theta_{\text{T},l}\right)\vect{u}_{\perp}\trans\left(\theta_{\text{T},l}\right) + \frac{J_{\tilde{\theta}_{\text{R},l}}}{d_{\text{R},\text{s},l}^2}\vect{u}_{\perp}\left(\theta_{\text{R},l} \right)\vect{u}_{\perp}\trans\left(\theta_{\text{R},l}\right).
			\IEEEeqnarraynumspace
		\end{IEEEeqnarray}
		In order to compute $\left(\vect{T}_{\text{np}} \vect{J}_{\vect{\phi}}\vect{T}_{\text{np}}\trans\right)^{-1}$ using block matrix inversion, we still need to compute $\vect{C}^{-1}$, which in turn, using block diagonal matrix inversion, requires the computation of $\vect{J}_{\text{s},l}^{-1}$. We compute $\vect{J}_{\text{s},l}^{-1}$ as~\cite[Lemma 1]{SW09}
		\begin{IEEEeqnarray}{rCl}
			\vect{J}_{\text{s},l}^{-1} &=& \frac{1}{\left|\vect{J}_{\text{s},l}\right|}\left[\frac{J_{\tau_l}'}{c^2}\vect{\nu}_{l,\perp}\vect{\nu}_{l,\perp}\trans + \frac{J_{\tilde{\theta}_{\text{T},l}}}{d_{\text{T},\text{s},l}^2}\vect{u}\left(\theta_{\text{T},l}\right)\vect{u}\trans\left(\theta_{\text{T},l}\right)+ \frac{J_{\tilde{\theta}_{\text{R},l}}}{d_{\text{R},\text{s},l}^2}\vect{u}\left(\theta_{\text{R},l}\right)\vect{u}\trans\left(\theta_{\text{R},l}\right)\right],
			\IEEEeqnarraynumspace
		\end{IEEEeqnarray}
		with $\vect{\nu}_{l,\perp} = \vect{u}_{\perp}\left(\theta_{\text{T},l}\right) + \vect{u}_{\perp}\left(\theta_{\text{R},l}\right)$ and
		\begin{IEEEeqnarray}{rCl}
			\left|\vect{J}_{s,l}\right| &=& \frac{J_{\tau_l}'}{c^2} \left(1 + \cos\left(\Delta\theta_l\right)\right)^2 \left( \frac{J_{\tilde{\theta}_{\text{T},l}}}{d_{\text{T},\text{s},l}^2} + \frac{J_{\tilde{\theta}_{\text{R},l}}}{d_{\text{R},\text{s},l}^2} \right) + \frac{J_{\tilde{\theta}_{\text{T},l}} J_{\tilde{\theta}_{\text{R},l}} }{d_{\text{T},\text{s},l}^2 d_{\text{R},\text{s},l}^2}\sin^2\left(\Delta\theta_l\right).\label{eq:J_sl}
			\IEEEeqnarraynumspace
		\end{IEEEeqnarray}
		Therefore, using \eqref{eq:T_np J_phi T_np}-\eqref{eq:J_sl}, $\vect{u}\trans\left(\theta_{\text{T},l}\right)\vect{u}\left(\theta_{\text{R},l}\right) = \cos\left(\Delta\theta_l\right)$ and $\vect{u}\trans\left(\theta_{\text{T},l}\right)\vect{u}_{\perp} \left(\theta_{\text{R},l}\right)  = \sin\left(\Delta\theta_l\right)$, after some algebraic manipulations we find that
		\begin{IEEEeqnarray}{rCl}
			&&\vect{T}_{\text{po}} \vect{J}_{\vect{\phi}} \vect{T}_{\text{np}}\trans \left(\vect{T}_{\text{np}} \vect{J}_{\vect{\phi}}\vect{T}_{\text{np}}\trans\right)^{-1}\vect{T}_{\text{np}} \vect{J}_{\vect{\phi}} \vect{T}_{\text{po}}\trans = \sum_{l=1}^{L-1}\Bigg[\frac{\left(J_{\tau_l}'\right)^2}{c^4} \frac{\left(1 + \cos\left(\Delta\theta_l\right)\right)^2}{\left|\vect{J}_{s,l}\right|} \left( \frac{J_{\tilde{\theta}_{\text{T},l}}}{d_{\text{T},\text{s},l}^2} + \frac{J_{\tilde{\theta}_{\text{R},l}}}{d_{\text{R},\text{s},l}^2} \right)\vect{z}_{\tau_l}\vect{z}_{\tau_l}\trans\nonumber\\
			&& \quad + \frac{J_{\tilde{\theta}_{\text{R},l}}^2}{d_{\text{R},\text{s},l}^4}\left( \frac{J_{\tau_l}'}{c^2} \frac{\left(1 + \cos\left(\Delta\theta_l\right)\right)^2}{\left|\vect{J}_{s,l}\right|} + \frac{J_{\tilde{\theta}_{\text{T},l}}}{d_{\text{T},\text{s},l}^2} \frac{\sin^2\left(\Delta\theta_l\right)}{\left|\vect{J}_{s,l}\right|} \right) \vect{z}_{\theta_{\text{R},l}}\vect{z}_{\theta_{\text{R},l}}\trans\nonumber\\
			&& \quad + \frac{J_{\tau_l}'J_{\tilde{\theta}_{\text{T},l}}J_{\tilde{\theta}_{\text{R},l}}}{c^2 d_{\text{T},\text{s},l}^2 d_{\text{R},\text{s},l}^2}\frac{\left(1 + \cos\left(\Delta\theta_l\right)\right)\sin\left(\Delta\theta_l\right)}{\left|\vect{J}_{s,l}\right|}\left(\vect{z}_{\tau_l}\vect{z}_{\theta_{\text{R},l}}\trans + \vect{z}_{\theta_{\text{R},l}}\vect{z}_{\tau_l}\trans\right)\Bigg] + \frac{1}{\hat{K}_{\epsilon_{\text{clk}}}}\hat{\vect{z}}_{\epsilon_{\text{clk}}}\hat{\vect{z}}_{\epsilon_{\text{clk}}}\trans,\label{eq:T_p J_phi T_np inv(T_np J_phi T_np) T_np J_phi T_p}
			\IEEEeqnarraynumspace
		\end{IEEEeqnarray}
		where
		\begin{IEEEeqnarray}{rCl}
			\hat{K}_{\epsilon_{\text{clk}}} &=& \frac{J_{\epsilon_{\text{clk}}}' + \frac{1}{\sigma_{\text{clk}}^2} - \sum_{l=1}^{L-1} J_{\tau_l}'}{c^2} + \sum_{l=1}^{L-1}\left[ \frac{J_{\tau_l}'J_{\tilde{\theta}_{\text{T},l}}J_{\tilde{\theta}_{\text{R},l}}}{c^2 d_{\text{T},\text{s},l}^2 d_{\text{R},\text{s},l}^2\left|\vect{J}_{s,l}\right|}\sin^2\left(\Delta\theta_l\right)\right],\nonumber\\
			\hat{\vect{z}}_{\epsilon_{\text{clk}}} &=&\sum_{l=1}^{L-1}\frac{J_{\tau_l}'J_{\tilde{\theta}_{\text{T},l}}J_{\tilde{\theta}_{\text{R},l}}}{c^2 d_{\text{T},\text{s},l}^2 d_{\text{R},\text{s},l}^2\left|\vect{J}_{s,l}\right|}\sin\left(\Delta\theta_l\right)\vect{z}_l.\nonumber
			\IEEEeqnarraynumspace
		\end{IEEEeqnarray}
		Combining \eqref{eq:T_p J_phi T_p} and \eqref{eq:T_p J_phi T_np inv(T_np J_phi T_np) T_np J_phi T_p}
		we obtain the desired result.





	\bibliographystyle{IEEEtran}
	\bibliography{IEEEabrv,PerformanceLimitsofSingleAnchormmWavePositioning}

\end{document}